\documentclass[12pt]{article}

\newcommand{\be}{\begin{equation}}
\newcommand{\ee}{\end{equation}}
\usepackage{amsmath}
\usepackage{amssymb}
\usepackage{latexsym}
\usepackage{epsfig}

\newcommand{\half}{{\textstyle {1\over 2}}}

\newcommand{\R}{\mathbb{R}}

\newcommand{\bea}{\begin{eqnarray}}
\newcommand{\eea}{\end{eqnarray}}

\newcommand{\RRR}{{\hbox{\rm R\kern-2.35mm R}}}

\def\ZZZ{{\hbox{ Z\kern-1.6mm Z}}}

\begin{document}

\begin{titlepage}
%\rightline\today 
\rightline{June 2010}
\rightline{\tt  Imperial-TP-2010-CH-03}
\rightline{\tt MIT-CTP-4154}
\begin{center}
\vskip 2.5cm
{\Large \bf {
Generalized metric formulation of double field theory}}\\
\vskip 2.0cm
{\large {Olaf Hohm${}^1\hskip-4pt ,$ Chris Hull${}^2\hskip-4pt ,$ and Barton Zwiebach${}^1$}}
\vskip 0.5cm
{\it {${}^1$Center for Theoretical Physics}}\\
{\it {Massachusetts Institute of Technology}}\\
{\it {Cambridge, MA 02139, USA}}\\
ohohm@mit.edu, zwiebach@mit.edu
\vskip 0.7cm
{\it {${}^2$The Blackett Laboratory}}\\
{\it {Imperial College London}}\\
{\it {Prince Consort Road}}\\
{\it { London SW7 @AZ, U.K.}}\\
c.hull@imperial.ac.uk

\vskip 2.5cm
{\bf Abstract}

\end{center}

\vskip 0.5cm

\noindent
\begin{narrower}

The generalized metric is a T-duality covariant symmetric matrix
constructed from~the metric and two-form gauge field and arises in generalized geometry. We view it here as a metric on the doubled spacetime and use it to give a simple formulation
with manifest T-duality  of the double field theory that describes
 the massless sector of closed strings. The gauge transformations are written in terms of a generalized Lie derivative
 whose commutator algebra is defined by a double field theory extension of the Courant bracket.

\end{narrower}

\end{titlepage}

\newpage

\tableofcontents
\baselineskip=16pt

\section{Introduction}

The remarkable T-duality properties of string theory~\cite{Giveon:1994fu} have motivated much study of field theory models that may incorporate
such properties.
Double field 
theory~\cite{Hull:2009mi,Hull:2009zb,Hohm:2010jy}  
 is a field theoretic approach inspired by closed string field 
theory~\cite{Kugo:1992md,Zwiebach:1992ie} that focuses
on the gravity, antisymmetric tensor, and dilaton fields.
These fields depend on
a doubled set of coordinates:  coordinates $x^i$ associated with
momentum excitations and coordinates $\tilde x_i$ associated
with winding excitations. 
The closed string
theory constraint $L_0-\bar L_0=0$ has implications:  the fields
and gauge parameters of doubled field theory must
be annihilated by the
differential operator $\partial_i \tilde \partial^i$, where a sum over~$i$ is
understood. Double field theory 
remains to be fully constructed; the work in~\cite{Hull:2009mi} gave
the doubled action only to cubic order in the fluctuations of fields
around a fixed background.
Noteworthy early work in double field theory includes that of
Tseytlin~\cite{Tseytlin:1990nb} and
Siegel~\cite{Siegel:1993th,Siegel:1993xq}.
Indeed, some of our results 
are closely related to the results of Siegel~\cite{Siegel:1993th,Siegel:1993xq}.

In a recent paper~\cite{Hohm:2010jy} we imposed a stronger 
form of the constraint $\partial_i \tilde \partial^i=0$ and constructed a manifestly background independent double field theory action for ${\cal E}_{ij}=g_{ij}+b_{ij}$, with $i,j = 1, 2, \ldots , D$,
and the dilaton $d$.  The action takes the
form:\footnote{Our notation can deal with a theory with
both compact and non-compact directions.
The spacetime has dimension $D= n+ d$
and is the product of $n$-dimensional Minkowski space
$\mathbb{R}^{n-1,1}$ and a torus $T^d$.  Although we write
$O(D,D)$ matrices,  the ones that are used describe T-dualities that belong to the
$O(d,d)$ subgroup associated with the torus.}

 \bea
 \label{THEActionINTRO}
 \begin{split}\hskip-10pt
  S \ = \ \int \,dx d\tilde{x}~
  e^{-2d}\Big[&
  -\frac{1}{4} \,g^{ik}g^{jl}   \,   {\cal D}^{p}{\cal E}_{kl}\,
  {\cal D}_{p}{\cal E}_{ij}
  +\frac{1}{4}g^{kl} \bigl( {\cal D}^{j}{\cal E}_{ik}
  {\cal D}^{i}{\cal E}_{jl}  + \bar{\cal D}^{j}{\cal E}_{ki}\,
  \bar{\cal D}^{i}{\cal E}_{lj} \bigr)~
\\ &    + \bigl( {\cal D}^{i}\hskip-1.5pt d~\bar{\cal D}^{j}{\cal E}_{ij}
 +\bar{{\cal D}}^{i}\hskip-1.5pt d~{\cal D}^{j}{\cal E}_{ji}\bigr)
 +4{\cal D}^{i}\hskip-1.5pt d \,{\cal D}_{i}d ~\Big]\;,
 \end{split}
 \eea
where the calligraphic derivatives ${\cal D}_i$ and $\bar{\cal D}_i$
are defined by
 \be
\label{groihffkdf}
{\cal D}_i \ \equiv \ {\partial\over \partial x^i} - {\cal E}_{ik} \,{\partial \over \partial\tilde x_k}\,,
~~~~\bar {\cal D}_i \ \equiv \ {\partial\over \partial x^i} + {\cal E}_{ki}\, {\partial \over \partial\tilde x_k}\,.
\ee
This action is T-duality invariant.  More precisely, it is invariant under the non-linear $O(D,D)$ transformations
 \begin{equation}\label{ODDtrans}
  {\cal E}^{\prime}(X^{\prime}) \ = \ (a{\cal E}(X)+b)(c{\cal E}(X)+d)^{-1}\;,
  \quad
  d^{\prime}(X^{\prime}) \ = \ d(X)\;, \quad X' = h X\,\,.  
 \end{equation}
 Here we have used matrix notation for the ${\cal E}$ field,  
  $a,b,c,d$ are the $D\times D$ blocks of an $O(D,D)$ matrix $h$,
  \begin{equation}\label{ODDelement}
 h= \begin{pmatrix} a &   b \\ c & d \end{pmatrix} \ \in \ O(D,D)\;,\quad
 h^t \eta h = \eta \quad
 \quad \hbox{with} \quad  \eta = \begin{pmatrix}
0&1 \\1&0 \end{pmatrix} \,,
 \end{equation}
and the coordinates have been grouped into the $O(D,D)$ vector
\be
X^M = \begin{pmatrix} \tilde x_i \\ x^i \end{pmatrix} \, , \qquad
\partial_M =   \begin{pmatrix} \tilde \partial^i  \\ \partial_i \end{pmatrix}\,.
\ee
$O(D,D)$ indices $M, N$ are raised and lowered with the {\em constant}
$O(D,D)$ invariant metric
\be
\label{etais}
\eta_{MN} =  \begin{pmatrix}
0&1 \\1&0 \end{pmatrix}\,.
\ee
If some of the coordinates are compact, the symmetry $O(D,D)$ is broken to the subgroup preserving the periodic boundary conditions.
Each term in the action~(\ref{THEActionINTRO}) is separately $O(D,D)$ invariant.  As explained in~\cite{Hohm:2010jy,Hull:2009mi} this result
largely follows  from consistent index contractions. Although we do not
display them explicitly, there are two
types of indices: unbarred and barred.  The first index in ${\cal E}_{ij}$ is
viewed as unbarred and the second index is viewed as barred. The index
in ${\cal D}_i$ is viewed as unbarred and the index in $\bar{\cal D}_i$
is viewed as barred.  Finally the indices in $g^{ij}$ can be viewed either as both unbarred
or as both barred.  Any term in which all contractions can be viewed as contractions of like-type indices is $O(D,D)$ invariant.\footnote{Also needed is that each ${\cal E}$ field appear with one calligraphic derivative. If more
than one derivative is used on a field, one must employ the $O(D,D)$ covariant derivatives discussed in~\cite{Hohm:2010jy}.}  While the $O(D,D)$
transformations are global, the various ingredients in the action $( g^{ij}, {\cal D} {\cal E}, \bar{\cal D} {\cal E}, {\cal D} d , \bar{\cal D} d$) transform
by the action of matrices that involve the field ${\cal E}$ and thus do not
define linear representations of the $O(D,D)$ group.  The barred/un-barred structure originates from the left/right factorization of closed string theory
and its geometric significance will be discussed in section 5.
The $O(D,D)$ symmetry is 
not manifest because the action
does not use conventional $O(D,D)$ tensors that
carry $O(D,D)$ indices
 ($M,N, \ldots= 1,2,\dots , 2D$).
 Since $\partial^M \partial_M = 2\partial_i \tilde \partial^i$, the constraint
on all fields and gauge parameters is $O(D,D)$ invariant.

\medskip
The action~(\ref{THEActionINTRO}) is also invariant under gauge transformations with a gauge parameter $\xi^M$:
\be
\xi^M = \begin{pmatrix} \tilde \xi_i \\[0.4ex] \xi^i \end{pmatrix} \, .
\ee
These parameters depend on both $x$ and $\tilde x$ coordinates.
  The gauge transformations
take the form
 \begin{equation}
 \label{finalgtINTRO}
 \begin{split}
  \delta {\cal E}_{ij} \ &= \ {\cal D}_i\tilde{\xi}_{j}-\bar{{\cal D}}_{j}\tilde{\xi}_{i}
  +\xi^{M}\partial_{M}{\cal E}_{ij}
  +{\cal D}_{i}\xi^{k}{\cal E}_{kj}+\bar{\cal D}_{j}\xi^{k}{\cal E}_{ik}\;,\\[0.5ex]
 \delta d~ \ &= - {1\over 2}  \partial_M \xi^M + \xi^M \partial_M \,d\,.
 \end{split}
 \end{equation}
Here $\xi^M\partial_M = \xi^i \partial_i + \tilde \xi_i \tilde \partial^i$ and
$\partial_M\xi^M = \partial_i\xi^i  +\tilde \partial^i \tilde \xi_i $.
The invariance of the action~(\ref{THEActionINTRO}) 
requires a strong version
of the constraint: $\partial^M \partial_M$ must
 annihilate all possible
{\em products} of fields and/or gauge parameters.
This constraint is so strong that  it
implies  that the theory is not
truly doubled: there is a choice of coordinates $(x', \tilde x')$, related to the original coordinates  $(x, \tilde x)$ by  $O(D,D)$, in which
the doubled fields do not depend on
the $\tilde x'$ coordinates~\cite{Hohm:2010jy}.
This means that we then have a field theory on the
subspace with coordinates $x'$ in which the gauge symmetry reduces to diffeomorphisms and $b$-field gauge transformations on that subspace.

Even though the theory 
it is not truly doubled,   the action~(\ref{THEActionINTRO}) is interesting because 
it exhibits new structures and
has some properties that are expected to persist 
 in the -- yet to be constructed -- general 
 double field theory.
 It is a natural action  for the 
field ${\cal E}_{ij}$ and inherits from string theory  a left-right structure that  is not present in the usual formulation. 
The gauge algebra is defined by the Courant bracket, or more precisely, an extension appropriate for doubled fields. Furthermore, the
action~(\ref{THEActionINTRO}), expanded to cubic order in fluctuations around a flat background
is fully gauge invariant to that order {\em without} imposing the strong
version of the constraint:
only the weak constraint is needed. We believe that the general
theory should be some natural generalization of the theory discussed here.

The gauge invariance of
the action~(\ref{THEActionINTRO}) is not manifest
and was verified in \cite{Hohm:2010jy}
through an elaborate and
 lengthy calculation. The above gauge transformations
can be rewritten in
suggestive ways but
remain mysterious. In this paper we
 provide an equivalent form of the action (\ref{THEActionINTRO}) for which the proof of gauge invariance is significantly simplified. Even the $O(D,D)$ invariance will be simpler:
all objects will transform in linear representations.

\medskip
The key object in the new construction will be the so-called ``generalized
metric".  This is a $2D\times 2D$ symmetric matrix
constructed from
the $D\times D$ metric tensor $g_{ij}$ and the antisymmetric tensor
$b_{ij}$ with the remarkable  property that it transforms
as an $O(D,D)$ tensor.
The explicit form of the  generalized metric is:\footnote{Note that this form follows from our convention $X^M=(\tilde x_i, x^i)$. Some  papers use the opposite conventions with  $X^M=( x^i,\tilde x_i)$, which would then lead to an expression for the generalized metric related to ours by swapping rows and columns.}
 \be
 \label{gennmet}
  {\cal H}_{MN} \  \ = \
  \begin{pmatrix}    g^{ij} & -g^{ik}b_{kj}\\[0.5ex]
  b_{ik}g^{kj} & g_{ij}-b_{ik}g^{kl}b_{lj}\end{pmatrix}\;.
 \ee
The non-linear $O(D,D)$ transformation
 (\ref{ODDtrans}) of the fields $g$ and $b$ implies a simple transformation
for $ {\cal H}_{MN}$.  Writing $X'= h X$ as $ X'^M = h^M{}_N X^N$ 
one finds:
 \be
h^P{}_Mh^Q{}_N {\cal H}'_{PQ}    (X')  \ = \  {\cal H}_{MN} (X)\,,
 \ee
 so that ${\cal H}_{MN}$ 
  is an $O(D,D)$ tensor, as indicated by the indices $M,N$.

 The matrix (\ref{gennmet})
appeared in the early
T-duality
literature. It defines the first-quantized Hamiltonian for closed strings in a toroidal background with constant metric and antisymmetric tensor fields~\cite{Giveon:1988tt,Duff:1989tf}.
Such matrices parameterize the coset space $O(D,D)/O(D)\times O(D)$,
and so arise in the toroidal  dimensional reduction of supergravity and string theories whose moduli take values in this coset \cite{Maharana:1992my,Meissner:1991zj,Kleinschmidt:2004dy}.

The doubled space then has two metrics, the constant $\eta_{MN}$ with signature $(D,D)$ and the metric
 ${\cal H}_{MN}$ which incorporates the dynamical fields and is positive definite if $g_{ij}$ is.
 Throughout this paper, we will always use the metric $\eta_{MN}$
 and its inverse $\eta^{MN}$  to lower and raise indices.
 Raising one or both indices with $\eta$
  defines
 the new tensors
 ${\cal H}^{MN}$ and ${\cal H}^M{}_{N}$.
 A striking feature of matrices of the form (\ref{gennmet})
 is that ${\cal H}^{MN}$ is the inverse of ${\cal H}_{MN}$:
 \be
 \label{inverseH}
 {\cal H}^{MP}{\cal H}_{PN}\ = \ \delta ^M{}_N\;.
 \ee
 Then ${\cal H}_{MN}$ can be viewed as a metric on the doubled space that satisfies 
the constraint that its inverse is
${\cal H}^{MN}\equiv
\eta ^{MP}{\cal H}_{PQ}\eta^{QN}$.
We define the matrix $S$ whose components $S^M{}_N$ 
are
 \be
 \label{definesmatrix}
S ^M{}_N\ \equiv \   
 {\cal H}^M{}_{N} \ = \    
 \eta ^{MP}{\cal H}_{PN} = {\cal H}^{MP} \eta_{PN}\,.
 \ee
The matrix $S$ satisfies
 \be
 \label{ssquared}
  S^2\ =\ 1\;,
 \ee
so that $S$ 
 is an almost local product structure, or almost real structure on the doubled space.
It has $D$ eigenvalues $+1$ and $D$ eigenvalues $-1$,
 and is an element of $O(D,D)$:
 \be
 \label{sinodd} 
 S^t\eta S\ =\ \eta\;.
 \ee

 In the mathematical literature, the generalized metric and the Courant bracket are key  structures in generalized geometry~\cite{Hitchin,Gualtieri,Tcourant}.
In this geometry the coordinates of the spacetime manifold $M$
are not doubled, rather, the tangent bundle $T$ of $M$ and the cotangent bundle $T^*$ of $M$ are put together to form a larger bundle $E= T\oplus T^*$ (or a twisted version  of this bundle).  Sections of this bundle $E$ are the formal sums
$X+ \xi$ of vectors $X$
and one-forms~$\xi$.  There is a natural
(indefinite) metric $\eta$ on sections of $E$ given by
$\langle X_1 + \xi_1 , X_2 + \xi_2\rangle = {X_1}^i \xi_{2i} + X_2^i \xi_{1i}$.
Introducing a metric $g$ and 2-form $b$ on $ M$ allows the definition of
tensor fields ${\cal H}$ and  $S$   by the formulae above.
The tensor $S$ defines a splitting $E = C_+\oplus C_-$ such
that $\langle \cdot , \cdot \rangle$ is positive definite on $C_+$
and negative definite on $C_-$.  The spaces $C_\pm$ are eigenspaces
of the  matrix  $S$ with eigenvalues $\pm 1$.
Gualtieri~\cite{Gualtieri} referred to $S$ as the generalized metric.
In \cite{Hull:2004in}, it was suggested that the term generalized metric be used instead for  ${\cal H}_{MN}$.
The generalized metric is then a $2D\times 2D$ matrix field  on   the $D$ dimensional space $M$ and a metric on sections of~$E$.   
In the context of  doubling, however, it was proposed  in~\cite{Hull:2004in} that the generalized metric be used as a conventional metric on the $2D$ dimensional doubled space. In this context the name  \lq generalized metric' is a misnomer and ${\cal H}_{MN}$ is better regarded as a conventional metric on the doubled space satisfying the constraint~(\ref{inverseH}). We will follow~\cite{Hull:2004in} and the subsequent literature and continue to refer to the metric
${\cal H}_{MN}$ on the doubled space as a generalized metric.

In this paper we present a double field theory spacetime action based on the generalized metric which is a rather nontrivial
and surprising rewriting of~(\ref{THEActionINTRO}).
The action is built using the $O(D,D)$ tensors ${\cal H}^{MN}$,
${\cal H}_{MN}$, and the derivatives~$\partial_M$ and is a 
rather  
simple and natural expression:
\bea\label{Hactionx}
\begin{split}
  S \ = \ \int dx d\tilde{x}\,e^{-2d}~\Big(~&\frac{1}{8}\,{\cal H}^{MN}\partial_{M}{\cal H}^{KL}
  \,\partial_{N}{\cal H}_{KL}-\frac{1}{2}{\cal H}^{MN}\partial_{N}{\cal H}^{KL}\,\partial_{L}
  {\cal H}_{MK}\\
  &-2\,\partial_{M}d\,\partial_{N}{\cal H}^{MN}+4{\cal H}^{MN}\,\partial_{M}d\,
  \partial_{N}d~\Big)\,.
 \end{split}
 \eea
This action is manifestly $O(D,D)$ invariant because all $O(D,D)$  indices are properly contracted.
The factor $e^{-2d}$ tranforms as  a density under gauge transformations and a scalar under $O(D,D)$ transformations.
Most directly, we view the above action as an
action for $g, b$ and $d$ fields, in which $g$ and $b$ enter through
${\cal H}$. With this identification the {\em Lagrangians}
associated with~(\ref{THEActionINTRO}) and (\ref{Hactionx}) are in fact identical. 
Alternatively, and more intriguingly,
one may view ${\cal H}$ as an elementary constrained
field with a natural geometric interpretation.

\medskip
The action~(\ref{Hactionx})  
 is gauge invariant provided the strong constraint is imposed.
The dilaton gauge
transformation in (\ref{finalgtINTRO}) is already in $O(D,D)$ covariant notation.
For ${\cal H}^{MN}$  
we find
 \be
  \delta_{\xi}{\cal H}^{MN} \ = \ \xi^{P}\partial_{P}{\cal H}^{MN}
+(\partial^{M}\xi_{P} -\partial_{P}\xi^{M})\,{\cal H}^{PN}
  +
 ( \partial^{N}\xi_{P} -\partial_{P}\xi^{N})\,{\cal H}^{MP}\;.
 \ee
This transformation looks like a diffeomorphism in which each index
gives a covariant and contravariant contribution.  We can view the
above right-hand side
as the generalized Lie derivative $\widehat{\cal L}_\xi$
of ${\cal H}^{MN}$ and write
\be
 \delta_{\xi}{\cal H}^{MN}  =  \widehat{\cal L}_\xi {\cal H}^{MN}\,.
\ee
We can indeed define  the action of $\widehat{\cal L}_\xi$ on an arbitrary {\em generalized tensor}
$A_{M_1M_2 \ldots}^{~N_1 N_2 \ldots}$
consistently with the derivation
property. The algebra of gauge transformations in the theory becomes the commutator algebra of
the generalized Lie derivatives. The commutator of generalized
Lie derivatives is in fact a generalized Lie derivative.
Indeed,
making use of the strong form of the constraint, we show that
\be
\bigl[ \,  \widehat{{\cal L}}_{\xi_1}\,,  \widehat{{\cal L}}_{\xi_2} \, \bigr]\,
= - \widehat{{\cal L}}_{[\xi_1, \xi_2]_{{}_{\rm C}}} \,,
 \ee
where the C bracket $[ \cdot \,, \cdot ]_{\rm{C}}$ is defined by
 \be
 \label{cbracketdef}
  \big[ \xi_1,\xi_2\big]_{\rm{C}}^{M} \ \equiv 
   \ \xi_{[1}^{N}\partial_{N}\xi_{2]}^M -\frac{1}{2}\,  \xi_{[1}^P\partial^{M}\xi_{2]\,P}\;,
 \ee
with $[ij] = ij - ji$.  The C~bracket, 
introduced by Siegel in~\cite{Siegel:1993th},
was recognized in~\cite{Hull:2009zb} as 
the $O(D,D)$ covariant extension of 
the Courant bracket for doubled fields.
As we shall discuss, our generalized Lie derivatives are closely related to those of~\cite{Siegel:1993th} and~\cite{Grana:2008yw}.

\section{$O(D,D)$ and the generalized metric}\setcounter{equation}{0}

In this section we summarize some well-known facts
about the $O(D,D)$ group, its Lie algebra, and the generalized
metric.  Some of these facts were already mentioned in the
introduction.

We define $O(D,D)$, as in  (\ref{ODDelement}),  as the group of $2D\times 2D $ matrices $h$
satisfying  
\be
\label{odddc}
h^t \eta\,  h \ =\  \eta\,, 
\ee and consequently
$h^{-1}  =  \eta ^{-1}h^t \eta \,.$
The associated Lie algebra generators $T$ satisfy
$T^t \eta + \eta T = 0$.
More explicitly,
\be
\label{liealg}
T = \begin{pmatrix} \alpha& \beta \\ \gamma & \delta \end{pmatrix}
\quad \to \quad  \gamma, \beta , ~\hbox{antisymmetric and}~ \delta
= - \alpha^t\,,
\ee
giving a total of $2D^2 -D$ parameters.

\bigskip
Raising the indices on the generalized metric~(\ref{gennmet})
gives  
 \be
 \label{huppindx}
 {\cal H}^{MN}= \eta^{MP}\eta^{NQ} \,{\cal H}_{PQ}\,.  
 \ee  
It will be convenient to use an index-free matrix notation.
 We will write  
 ${\cal H}$ to denote the matrix whose components   
  are ${\cal H}^{MN}$ : 
   \be
   \label{hmatrixd}
  {\cal H}~ \equiv ~ {\cal H}^{\bullet\,\bullet}  \,,
  \ee 
  with the heavy dots indicating the index positions.   
We write $\eta$ to denote the matrix whose
components are $\eta_{MN}$ and $S$ 
to denote the matrix
whose components are $S^M{}_N$:
\be
\eta\  \equiv\  \eta_{\bullet\,\bullet} \,, ~~~~~  S 
\ \equiv \   S^\bullet{}_\bullet \,.
\ee
In this notation (\ref{definesmatrix}) is written as
\be
\label{sasmatrix}
S \ = \ {\cal H} \,\eta \,.
\ee
  It  follows from (\ref{huppindx}),  (\ref{hmatrixd}), and (\ref{gennmet})  that
\be
\label{Hg}
  {\cal H} \ = \
  \begin{pmatrix}    g-bg^{-1}b & bg^{-1}\\[0.5ex]
  -g^{-1}b & g^{-1}\end{pmatrix}\, .
 \ee
The matrix ${\cal H}$  is  symmetric  ($ {\cal H}^t= {\cal H}$) 
and satisfies
 \be
 \label{propty}
 {\cal H}\,\eta \,{\cal H}=\eta^{-1}\;,
  \ee
so that its inverse is
\be
\label{hinv}
{\cal H}^{-1} = \eta {\cal H}^t \eta =  \eta {\cal H} \eta\,.
\ee
Then ${\cal H}$, with components ${\cal H}^{MN}$, is the inverse of the generalized metric 
with components ${\cal H}_{MN}$, which we denote ${\cal H}^{-1} $ so that (\ref{inverseH}) becomes
${\cal H}{\cal H}^{-1} =1$.
 One can check explicitly that the matrix in (\ref{Hg})  is indeed the inverse of the matrix in (\ref{gennmet}).

Since the entries of the matrix $\eta^{-1}$ coincide with those of the
matrix $\eta$ and ${\cal H}$ is symmetric, 
${\cal H}$ satisfies the defining condition~(\ref{odddc}) for $O(D,D)$.
We will then refer to a symmetric matrix ${\cal H}$ that satisfies the constraint (\ref{propty})
as being a (symmetric) $O(D,D)$ matrix.  Strictly speaking, this is an abuse of language as $\cal H$ has upper indices ${\cal H}^{MN}$ while group elements have mixed indices
$h^M{}_N$.
 The fact that $\cal H$ is a symmetric $O(D,D)$ matrix can be seen
 explicitly  by writing ${\cal H}$ as the product of three simple $O(D,D)$ matrices:
  \be
  {\cal H} \ =\
  \begin{pmatrix}
  g-bg^{-1}b   & bg^{-1}\\[0.5ex] -g^{-1}b & g^{-1}
  \end{pmatrix} =
  \begin{pmatrix}
  1   & b\\[0.5ex] 0 & 1
  \end{pmatrix}
    \begin{pmatrix}
  g   & 0\\[0.5ex] 0 & g^{-1}
  \end{pmatrix}
 \begin{pmatrix}
    1  & 0\\[0.5ex] -b & 1
  \end{pmatrix}\,.
 \ee
The construction of ${\cal H}$ and its remaining properties are motivated
by the action~(\ref{ODDtrans}) of $h\in O(D,D)$ on ${\cal E}$.  We have
$
{\cal E}' =  h ({\cal E}) =  (a {\cal E} + b ) (c{\cal E} + d)^{-1}$.
Let $h_{\cal E}$ be the $O(D,D)$  transformation that
 creates ${\cal E}$ starting from the identity background\footnote{For simplicity we give the argument for Euclidean signature. For Lorentzian signature, we take ${\cal I}$  as the Minkowski metric and
 $g = e{\cal I}e^t$.}
 ${\cal E}={\cal I} $, where ${\cal I}$ is the unit matrix
 ${\cal I}_{ij}=\delta _{ij}$, so that it satisfies
${\cal E} =  h_{\cal E} ({\cal I})$.
Such a transformation is
 given by
\be\label{representative}
 h_{\cal E} = \begin{pmatrix} e  & b (e^t)^{-1} \\[1.0ex] 0 & (e^t)^{-1} \end{pmatrix}\;,
 \ee
where we have introduced a vielbein $e $ for the metric, so that
\be
g = ee^t \, .
\ee
 Indeed
we easily confirm that, as desired,
\be
h_{\cal E} ({\cal I})  = ( e {\cal I} + b (e^t)^{-1} ) ( 0\cdot {\cal I} + (e^t)^{-1})^{-1} =  ( e + b (e^t)^{-1} )  e^t =  e e^t  + b = g+ b = {\cal E}\,.
\ee
The matrix $h_{\cal E}$ is not uniquely defined;  right multiplication by the $O(D)\times O(D)$ subgroup of $O(D,D)$ that leaves the background ${\cal I}$ invariant gives another transformation with the desired properties.

 Let us now consider the group action on $h_{\cal E}$.
 Consider an $O(D,D)$ transformation $h$  taking ${\cal E}$ to
${\cal E}' =  h ({\cal E})$.
We then have
$
h_{{\cal E}'} ({\cal I} ) =  {\cal E}' = h ( h_{\cal E} ({\cal I})) = ( h h_{\cal E}) ({\cal I})$.
We thus deduce that
\be
\label{trans}
h:  ~h_{\cal E} \to h_{{\cal E}'}=  h h_{\cal E}  \,.
\ee
We now define the {\em symmetric} $O(D,D)$  matrix
\be
\label{smnvm}
{\cal H}({\cal E}) \equiv  h_{\cal E} h_{\cal E}^t\,.
\ee
Here ${\cal H}$ is in
$O(D,D)$ because both $h_{\cal E}$ and $h_{\cal E}^t$ are.
Moreover, the $O(D)\times O(D)$ ambiguity in $h_{\cal E}$ drops out of ${\cal H}$.
 A quick computation
 confirms that this agrees with~(\ref{Hg}):  
 \begin{equation*}
{\cal H}({\cal E}) = \begin{pmatrix} e  & b (e^t)^{-1} \\[1.0ex] 0 & (e^t)^{-1} \end{pmatrix}
\begin{pmatrix} e^t  & 0 \\[1.0ex] -e^{-1}b & e^{-1} \end{pmatrix}
= \begin{pmatrix}  ee^t - b (ee^t)^{-1}b & b (ee^t)^{-1}
\\[1.0ex] - (ee^t)^{-1}  b & (ee^t)^{-1}  \end{pmatrix}
= \begin{pmatrix}  g - bg^{-1}b & b g^{-1}
\\[1.0ex] - g^{-1}  b & g^{-1}  \end{pmatrix}\,.
\end{equation*}
It follows from (\ref{trans}) and (\ref{smnvm})  that under a transformation $h\in O(D,D)$
such that ${\cal E}'  = h{\cal E}$ we get the following transformation of
${\cal H}$:  
\be
\label{HTr}
{\cal H} ({\cal E}') =  h  \, {\cal H} ({\cal E}) \, h^t \,.
\ee

We note that ${\cal H}$ at  any 
point $X$ is a symmetric $O(D,D)$ matrix 
(i.e. a symmetric matrix satisfying (\ref{propty}))
that is in the component of the
$O(D,D)$ group connected to the identity.
The formula for ${\cal H}$ in terms of $g$ and $b$ is 
a useful parameterization of this symmetric $O(D,D)$ matrix.
We can readily check that the counting of degrees of freedom
works out.  This is most easily done near the identity, using
the Lie algebra results.    If $h = 1 + \epsilon T$ is to be symmetric then
we get that $\alpha$ is symmetric and $\gamma = -\beta$, referring to
the notation in (\ref{liealg}).  Thus the whole $T$ is characterized by
a symmetric $\alpha$ and an antisymmetric $\beta$.  This is precisely
$D^2$ parameters, the same number of parameter as in ${\cal E}$.

The $O(D,D)$ indices make the transformation properties manifest.
  For an $O(D,D)$ vector
$V^M$ and an $O(D,D)$ element $h$ we have a transformation
\be
{V'}^M  =  h^M{}_{N} \, V^N \,.
\ee
An upper index $M$ runs over $2D$ values, the first $D$ of them
described with a lower roman index $i$ and the second $D$ of them
with an upper roman index $i$:
\be
V^M  =  ( \, v_i\,, \, v^i\,) \,.
\ee
The indices $i,j=1,...,D$ label representations of the $GL(D,\R)$ subgroup of $O(D,D)$.
The components $v_i$ and $v^i$ are
independent.
For lower $O(D,D)$ indices we have
\be
U_M  =  ( \, u^i\,, \, u_i\,) \,.
\ee
In summary we can write ${}^{M} \ = \ \big(\,{}_{i}\,,\, {}^{i}\,\big)$
and ${}_{M} \ = \ \big(\,{}^{i}\,,\, {}_{i}\,\big)$.  The matrix $h^M{}_{N}$,
for example, has components
 \be
  h^M{}_{N} \ = \ \begin{pmatrix}     h_i{}^{j} & h_{ij}\\[0.5ex]
 h^{ij} & h^i{}_{j} \end{pmatrix} \,.
 \ee

We can view ${\cal H}$ as a tensor ${\cal H}^{MN}$
with two upper $O(D,D)$ indices
because the  transformation (\ref{HTr}) 
implies that:
\be
{{\cal H}'}^{MN} (X') =  h^M{}_{P} \, h^N{}_{Q}\, {\cal H}^{PQ}(X)\,.
\ee
We  thus identify the ${\cal H}$ matrix
with ${\cal H}^{MN} (= {\cal H}^{NM})$ as
 \bea\label{H}
  {\cal H}^{MN} \ = \ \begin{pmatrix}     {\cal H}_{ij} & {\cal H}_{i}{}^{j}\\[0.5ex]
  {\cal H}^{i}{}_{j} & {\cal H}^{ij} \end{pmatrix} \ = \
  \begin{pmatrix}    g_{ij}-b_{ik}g^{kl}b_{lj} & b_{ik}g^{kj}\\[0.5ex]
  -g^{ik}b_{kj} & g^{ij}\end{pmatrix}\;.
 \eea
  The   symmetry ${\cal H}^{MN} = {\cal H}^{NM}$
     implies that
\be
{\cal H}_{ij} = {\cal H}_{ji} \,, ~~{\cal H}^{ij} = {\cal H}^{ji}\,, ~~
{\cal H}_i^{~j} = {\cal H}^j_{~i}\,.
\ee
The relation  $\eta{\cal H}\eta= {\cal H}^{-1}$ in (\ref{hinv}) with $O(D,D)$ indices is
\be
 ({\cal H}^{-1})_{MN}
 =\eta_{MP} {\cal H}^{PQ} \eta_{QN}  = {\cal H}_{MN}\;,
\ee
so that ${\cal H}_{MN}$ is indeed the inverse of ${\cal H}^{MN}$.
The generalized metric is 
 \be
  {\cal H}_{MN} \ = \ \left({\cal H}^{MN}\right)^{-1} \ = \
  \begin{pmatrix}    g^{ij} & -g^{ik}b_{kj}\\[0.5ex]
  b_{ik}g^{kj} & g_{ij}-b_{ik}g^{kl}b_{lj}\end{pmatrix}\;.
 \ee

\section{Gauge symmetry and Courant brackets}\setcounter{equation}{0}
In this section we will show that the gauge transformations (\ref{finalgtINTRO}), which are non-linear when written in terms
of ${\cal E}_{ij}$, act linearly on the $O(D,D)$ covariant matrix ${\cal H}^{MN}$ introduced above. This linear form of the
gauge transformations naturally suggests a notion of generalized Lie derivative, for which a tensor calculus can be developed.
This simplifies the proof of gauge invariance to be undertaken in the next section. Finally, the closure of the gauge algebra
according to the Courant bracket will be checked in this formulation.

\subsection{Gauge transformations of the generalized metric} 
The gauge transformations (\ref{finalgtINTRO}) take a highly non-linear form when written in terms of the fundamental fields ${\cal E}_{ij}=g_{ij}+b_{ij}$.  
Writing out~$\delta{\cal E}_{ij}$  using the definition~(\ref{groihffkdf})
of calligraphic derivatives one determines the transformation of
$\delta_\xi g_{ij}$ (from which $\delta g^{ij}$ follows) 
and the transformation $\delta_\xi b_{ij}$.  The results are
 \be
 \label{nonlingauge}
 \begin{split}
  \delta_{\xi}g_{ij} &= {\cal L}_{\xi}g_{ij}+{\cal L}_{\tilde{\xi}}g_{ij}
     + \big(\tilde{\partial}^{k}\xi^{l}-\tilde{\partial}^{l}
   \xi^{k}\big)  (g_{ki}\,b_{jl} + g_{kj}\,b_{il} )\;, \\[0.8ex]
  \delta_{\xi}g^{ij} &= {\cal L}_{\xi}g^{ij}+{\cal L}_{\tilde{\xi}}g^{ij}
  -\big[\big(\tilde{\partial}^{i}\xi^{k}
  -\tilde{\partial}^{k}\xi^{i}\big)g^{jl}b_{lk}+\left(i\leftrightarrow j\right)\big]\;,
  \\[0.8ex]
  \delta_{\xi}b_{ij} &= {\cal L}_{\xi}b_{ij}+{\cal L}_{\tilde{\xi}}b_{ij}
  +\partial_{i}\tilde{\xi}_{j}-\partial_{j}\tilde{\xi}_{i}
  +g_{ik}\big(\tilde{\partial}^{l}\xi^{k}-\tilde{\partial}^{k}
  \xi^{l}\big)g_{lj}+b_{ik}\big(\tilde{\partial}^{l}\xi^{k}-\tilde{\partial}^{k}
  \xi^{l}\big)b_{lj}\;.
  \end{split}
 \ee
Here we use the Lie derivatives with respect to $\xi^{i}$ and 
dual Lie derivatives with respect to $\tilde{\xi}_{i}$. 
Their definition on tensors with arbitrary number of upper and
lower indices follow from 
 \be
 \begin{split}
  {\cal L}_{\xi}\,{u}_{i}^{~j}~&= ~\xi^{p}\partial_{p}{u}_{i}^{~j}
  +\partial_{i}\xi^{p}\,u^{~j}_{p}
  -\partial_{p}\xi^{j}\,u^{~p}_{i}\;, \\[1.0ex]
  {\cal L}_{\tilde{\xi}}\,{u}_{i}^{~j}~&= ~\tilde{\xi}_{p}\tilde{\partial}^{p}
  \hskip1pt{u}_{i}^{~j}
  +\tilde{\partial}^{j}\tilde{\xi}_{p}\,u^{~p}_{i}
  -\tilde{\partial}^{p}\tilde{\xi}_{i} \, u^{~j}_{p}\;.
\end{split}
\ee

It is of interest to determine the gauge transformations of the
(inverse)
generalized metric~${\cal H}^{MN}$. The direct computation
gives a remarkable result: the gauge
transformations of~${\cal H}^{MN}$
implied by (\ref{nonlingauge})
are linear in ${\cal H}^{MN}$.  
Indeed, we find
 \be
 \label{lingenmet}
 \begin{split}
  \delta{\cal H}_{ij} &= {\cal L}_{\xi}{\cal H}_{ij}+{\cal L}_{\tilde{\xi}}{\cal H}_{ij}
  +\big[(\partial_{i}\tilde{\xi}_{p}-\partial_{p}\tilde{\xi}_{i}){\cal H}^{p}{}_{j}
  +\left(i\leftrightarrow j\right)\big]\;, \\[0.8ex]
  \delta{\cal H}^{ij} &= {\cal L}_{\xi}{\cal H}^{ij}+{\cal L}_{\tilde{\xi}}{\cal H}^{ij}
  +\big[(\tilde{\partial}^{i}\xi^{p}-\tilde{\partial}^{p}\xi^{i}){\cal H}^{j}{}_{p}
  +\left(i\leftrightarrow j\right)\big]\;, \\[0.8ex]
  \delta{\cal H}^{i}{}_{j} &=
  {\cal L}_{\xi}{\cal H}^{i}{}_{j}+{\cal L}_{\tilde{\xi}}{\cal H}^{i}{}_{j}
  +(\tilde{\partial}^{i}\xi^{p}-\tilde{\partial}^{p}\xi^{i}){\cal H}_{pj}
  +(\partial_{j}\tilde{\xi}_{p}-\partial_{p}\tilde{\xi}_{j}){\cal H}^{ip}\;.
\end{split}
\ee
We sketch the proof of the first relation in (\ref{lingenmet}). For this we 
rewrite~(\ref{nonlingauge}) with a separation of terms
$\hat \delta_\xi$ that are quadratic in the fields: 
\be
 \label{nonlingauge99}
 \begin{split}
  \delta_{\xi}g_{ij} &= {\cal L}_{\xi}g_{ij}+{\cal L}_{\tilde{\xi}}g_{ij}
     + \hat\delta_\xi g_{ij} \;, \\[0.8ex]
  \delta_{\xi}g^{ij} &= {\cal L}_{\xi}g^{ij}+{\cal L}_{\tilde{\xi}}g^{ij}
+ \hat\delta_\xi g^{ij} \;,
  \\[0.8ex]
  \delta_{\xi}b_{ij} &= {\cal L}_{\xi}b_{ij}+{\cal L}_{\tilde{\xi}}b_{ij}
  +\partial_{i}\tilde{\xi}_{j}-\partial_{j}\tilde{\xi}_{i}
  +\hat\delta_\xi b_{ij} \;.
  \end{split}
 \ee
 The expressions for $\hat\delta_\xi$ on the fields follow directly
 by comparison with~(\ref{nonlingauge}).
In the computation~of 
\be
 \delta_{\xi}{\cal H}_{ij} = \delta_{\xi}\big(g_{ij}-b_{ik}g^{kl}b_{lj}\big)\,,
\ee
the terms in the gauge variations of fields that consist of Lie derivatives
combine to form the Lie derivatives of ${\cal H}_{ij}$.  We thus find
 \be
 \begin{split}
  \delta_{\xi}{\cal H}_{ij} 
   ~=&~ {\cal L}_{\xi}{\cal H}_{ij}+{\cal L}_{\tilde{\xi}}{\cal H}_{ij}
  -(\partial_{i}\tilde{\xi}_{k}-\partial_{k}\tilde{\xi}_{i})g^{kl}b_{lj}
  -b_{ik}g^{kl}(\partial_{l}\tilde{\xi}_{j}-\partial_{j}\tilde{\xi}_{l})
  +\hat{\delta}_{\xi}{\cal H}_{ij}\\[1.0ex]
  ~=&~  {\cal L}_{\xi}{\cal H}_{ij}+{\cal L}_{\tilde{\xi}}{\cal H}_{ij}
  +(\partial_{i}\tilde{\xi}_{p}-\partial_{p}\tilde{\xi}_{i}){\cal H}^{p}{}_{j}
  +(\partial_{j}\tilde{\xi}_{p}-\partial_{p}\tilde{\xi}_{j}){\cal H}^{p}{}_{i}
  +\hat{\delta}_{\xi}{\cal H}_{ij}\;,
 \end{split}
 \ee
where we have used~(\ref{H}) to identify components of
${\cal H}^{MN}$  and have  relabeled the indices. 
A direct computation then shows that:
 \be
  \hat{\delta}_{\xi}{\cal H}_{ij} \ = \ \hat{\delta}_{\xi}\big(g_{ij}-b_{ik}g^{kl}b_{lj}\big) \ = \ 0\;.
 \ee
This completes the proof that the gauge transformation is linear in 
${\cal H}$.
The other relations in (\ref{lingenmet}) follow similarly. 
 The linear part of the above computation essentially coincides 
with the analysis of \cite{Grana:2008yw}, but the remarkable cancellation 
of the non-linear terms is only visible once the 
dual derivatives~$\tilde{\partial}^{i}$ enter.

The transformations (\ref{lingenmet})
can be written in a manifestly $O(D,D)$ covariant form and the
result is rather simple:
 \be
 \begin{split}
  \delta_{\xi}{\cal H}^{MN} &=  \xi^{P}\partial_{P}{\cal H}^{MN}
  -\partial_{P}\xi^{M}\,{\cal H}^{PN}-\partial_{P}\xi^{N}\,{\cal H}^{MP}\\[0.5ex]
  &~~~+\eta_{PQ}\left(\eta^{MK}\partial_{K}\xi^{P}\,{\cal H}^{QN}+
  \eta^{NK}\partial_{K}\xi^{P}\,{\cal H}^{MQ}\right)\;.
 \end{split}
 \ee
The first terms are the standard diffeomorphism terms, while the remaining ones are novel and
responsible for closure into the
C~bracket. 
If we use the notation
$\partial^{M}=\eta^{MN}\partial_{N}$, $\xi_{M}=\eta_{MN}\xi^{N}$, etc., these gauge transformations can be rewritten in a even more suggestive form as
 \bea\label{finalgauge}
  \delta_{\xi}{\cal H}^{MN} \ = \ \xi^{P}\partial_{P}{\cal H}^{MN}
  -\partial_{P}\xi^{M}\,{\cal H}^{PN}-\partial_{P}\xi^{N}\,{\cal H}^{MP}
  +\partial^{M}\xi_{P}\,{\cal H}^{PN}+
  \partial^{N}\xi_{P}\,{\cal H}^{MP}\;. 
 \eea
It looks like a diffeomorphism which democratically treats the indices both as covariant and contravariant and can be seen as a generalized Lie derivative which we
consider in more detail in the next subsection.
Another convenient rewriting that groups the covariant and contravariant
action on each index is
 \be\label{finalgauge99}
  \delta_{\xi}{\cal H}^{MN} \ = \ \xi^{P}\partial_{P}{\cal H}^{MN}
 +(\partial^{M}\xi_{P}\, -\partial_{P}\xi^{M})         {\cal H}^{PN}+
 ( \partial^{N}\xi_{P}\,  -\partial_{P}\xi^{N}   ){\cal H}^{MP}\;.
 \ee
The gauge invariance has the usual gauge invariance:  gauge parameters
of the form $\xi^P  = \partial^P \chi$ generate no gauge
transformations:  $\delta_{\partial \chi} {\cal H}^{MN} = 0$.
This is readily verified in the equation above using the strong form
of the constraint.

\subsection{Generalized Lie derivatives and Courant brackets}

The transformation of ${\cal H}^{MN}$
in~(\ref{finalgauge99}) involves an operation similar to a Lie derivative. This motivates the definition of
a  {\em generalized} Lie derivative $\widehat{{\cal L}}$
of
 a {\em generalized}
 tensor which has  upper and lower indices $A_{N_1\ldots}^{~M_1 \ldots}$.
 For a tensor $A_M{}^N$ the
 {\em generalized} Lie derivative is defined to be
 \be
 \label{genLie}
  \widehat{{\cal L}}_{\xi} A_{M}{}^{N} \ \equiv \
  \xi^{P}\partial_{P}A_{M}{}^{N}
+ (\partial_{M}\xi^{P}-\partial^{P}\xi_{M})
  \,A_{P}{}^{N}
  +(\partial^{N}\xi_{P}\,    -\partial_{P}\xi^{N})       A_{M}{}^{P}\,.
 \ee
For multiple indices the generalized Lie derivative is defined analogously:
each index gives rise to two terms.
With such definition we immediately recognize
that the gauge transformation (\ref{finalgauge99}) of the generalized
(inverse) metric  is simply a generalized Lie derivative:
\be
\delta_{\xi}{\cal H}^{MN}=\widehat{{\cal L}}_{\xi}{\cal H}^{MN}\,.
\ee
The generalized Lie derivative differs from 
the conventional Lie derivative by terms 
that involve explicitly  the $O(D,D)$ metric  
 \be   \label{genLieLie}
  \begin{split}
  \widehat{{\cal L}}_{\xi} A_{M}{}^{N} \ &= \
  {{\cal L}}_{\xi} A_{M}{}^{N}
-\partial^{P}\xi_{M}
  \,A_{P}{}^{N}
  +\partial^{N}\xi_{P}\,         A_{M}{}^{P}
  \\  \ &= \
  {{\cal L}}_{\xi} A_{M}{}^{N}
-\eta ^{PQ}\eta_{MR} \ \partial_{Q}\xi^{R}
  \,A_{P}{}^{N}
  +  \eta _{PQ}\eta^{NR}\  \partial_{R}\xi^{Q}\,         A_{M}{}^{P}
  \,.
  \end{split}
\ee
With the  definition~(\ref{genLie})   
  $\widehat{\cal L}_{\xi}$ is   a derivative satisfying the Leibniz rule,
 \bea
  \widehat{\cal L}_{\xi}\left(A_{N_1\ldots}^{~M_1 \ldots} \,
  B_{P_1\ldots}^{~Q_1 \ldots} \right) \ = \
  \big(\widehat{\cal L}_{\xi}A_{N_1\ldots}^{~M_1 \ldots}\big)
  B_{P_1\ldots}^{~Q_1 \ldots}+
  A_{N_1\ldots}^{~M_1 \ldots}\big(\widehat{\cal L}_{\xi}
  B_{P_1\ldots}^{~Q_1 \ldots}\big)\;,
 \eea
 so that it is consistent to regard
 products of generalized tensors are generalized
tensors with the index structure of the full set of indices.
The generalized Lie derivative
$ \widehat{{\cal L}}_{\xi} $ of {\em any} generalized tensor
vanishes when $\xi^M = \partial^M \chi$, so that for
any generalized tensor $A$ we have  
\be
\widehat{\cal L}_{\xi+ \eta^{-1} \partial\chi}A= \widehat{\cal L}_{\xi}A\;.
\ee

A remarkable and important  property is that the  
generalized Lie 
derivatives of the $O(D,D)$ metric $\eta_{MN}$
and the Kronecker tensor $\delta_M{}^N$ vanish:
\be
 \widehat{\cal L}_{\xi}\eta_{MN}=0,\qquad
 \widehat{\cal L}_{\xi}\eta^{MN}=0 ,\qquad
  \widehat{\cal L}_{\xi}\delta _M{}^N=0\;.
  \ee
  For example,
 \be
  \widehat{\cal L}_{\xi}\eta^{MN} \ = \ \xi^{P}\partial_{P}\,\eta^{MN}-\partial^{N}\xi^{M}
  -\partial^{M}\xi^{N}+\partial^{N}\xi^{M}+\partial^{M}\xi^{N} \ = \ 0\;.
 \ee
 This is unusual; in ordinary diffeomorphism invariant theories a constant world-tensor with two covariant or two contravariant  indices does not have vanishing Lie derivative
along arbitrary vector fields.

An important consequence of $ \widehat{\cal L}_{\xi}\eta = 
\widehat{\cal L}_{\xi}\eta^{-1} =0$  
  is that 
the constraint that ${\cal H}$ is an $O(D,D)$ matrix  
 is compatible with its gauge symmetry. 
Taking the generalized 
Lie derivative of the condition 
$  {\cal H}\eta{\cal H} \ = \ \eta^{-1}$ gives
\be
    \big(\widehat{\cal L}_{\xi}{\cal H}\big)\eta{\cal H} 
  +{\cal H}\eta  \big(\widehat{\cal L}_{\xi}{\cal H}\big)  \ = \ 0\;.
 \ee
This means that    
\be
\big( \delta_{\xi}{\cal H}\big)\eta{\cal H} 
  +{\cal H}\eta
  \big(\delta_{\xi}{\cal H}\big)
  \ = \ 0\;,
\ee
so that ${\cal H}\eta{\cal H} \ = \ \eta^{-1}$   
  is preserved by the gauge transformations, showing that  the $O(D,D)$ and gauge symmetries are compatible. 

The constant tensors $\eta $ and $\delta$ can be used to simplify and relate tensor expressions.
The simplest generalized tensor is a scalar $S$ for which
\be
\widehat{\cal L}_\xi  S =  \xi^P \partial_P S \,.
\ee
A generalized tensor with two indices contracted, such as $A_M{}^{M}$,
is a generalized scalar (the definition~(\ref{genLie})
gives $\widehat{\cal L}_{\xi}A_M{}^{M} = \xi^P \partial_P A_M{}^{M}$).
Any contraction of an upper and a lower index effectively removes
both indices from the tensor.
For a tensor $A_M$ with one index down we have
 \be\label{genLievec}
  \widehat{{\cal L}}_{\xi} A_{M} \ = \ \xi^{P}\partial_{P}A_{M}{}
  +(\partial_{M}\xi^{P}
  \, -\partial^{P}\xi_{M})\,A_{P}\;.
 \ee
 For a tensor $A^M$ with one index up we have
 \be\label{genLievecup}
  \widehat{{\cal L}}_{\xi} A^{M} \ = \ \xi^{P}\partial_{P}A^{M}{}
  +(\partial^{M}\xi_{P}
  \, -\partial_{P}\xi^{M})\,A^{P}\;,
 \ee
 so that $\widehat{{\cal L}}_{\xi} A^{M}  = \eta^{MN} \widehat{{\cal L}}_{\xi} A_{M}$ as expected.

\bigskip
The algebra of generalized Lie derivatives is governed by
the C~bracket~(\ref{cbracketdef}).
The commutator algebra
of generalized Lie derivatives is most easily calculated
acting on the generalized tensor $A_M$.
A straightforward computation gives
\be
\bigl[ \,  \widehat{{\cal L}}_{\xi_1}\,,  \widehat{{\cal L}}_{\xi_2} \, \bigr]\, A_M
= - \widehat{{\cal L}}_{[\xi_1, \xi_2]_{{}_{\rm C}}}  A_M
+ F_M (\xi_1, \xi_2, A)\,,
\ee
where  $[ \cdot \,, \cdot ]_{\rm{C}}$ is the C~bracket defined
in~(\ref{cbracketdef}) and
$F_M$ is  given by
\be
F_M (\xi_1, \xi_2, A)=
- {1\over 2}  \xi_{[1 N} \,\partial^Q \xi_{2]}^N \, \partial_Q A_M
+ \partial^Q  \xi_{[1M} \, \partial_Q \xi_{2]}^P  \,  A_P\,,
\ee
which vanishes by the strong form of the constraint.
Since we always assume this constraint, we have shown that acting
on a field $A_M$ we have
\be
\label{liegencomm}
\bigl[ \,  \widehat{{\cal L}}_{\xi_1}\,,  \widehat{{\cal L}}_{\xi_2} \, \bigr]
= - \widehat{{\cal L}}_{[\xi_1, \xi_2]_{{}_{\rm C}}}\,.
\ee

This commutator actually holds acting on arbitrary
generalized tensors.  Indeed, consider the action on the product
of two one-index generalized tensors
 \be
 \begin{split}
  \big[\, \widehat{\cal L}_{\xi_1},\widehat{\cal L}_{\xi_2}\,\big]\big(A_M B_N\big) \ &= \
  \big(\big[\, \widehat{\cal L}_{\xi_1},\widehat{\cal L}_{\xi_2}\,\big]A_M\big) B_N +
  A_{M}\big[\, \widehat{\cal L}_{\xi_1},\widehat{\cal L}_{\xi_2}\,\big]B_N
  \\[0.8ex]
    \ &= \ -\widehat{\cal L}_{[\xi_1,\xi_2]_{C}}\big(A_M B_N\big).
 \end{split}
 \ee
By iterating this proof it follows that the commutator property
(\ref{liegencomm}) holds for all tensors with lower indices. It also holds for tensors with an arbitrary number of upper indices. This follows from
$\widehat{\cal L}_{\xi}\eta^{MN}=0$,
 \be
 \begin{split}
  \big[\, \widehat{\cal L}_{\xi_1},\widehat{\cal L}_{\xi_2}\,\big]A^{M} &=
  \big[\, \widehat{\cal L}_{\xi_1},\widehat{\cal L}_{\xi_2}\,\big]\big(\eta^{MN}A_{N}\big)
  \ = \
  \eta^{MN}\big[\, \widehat{\cal L}_{\xi_1},\widehat{\cal L}_{\xi_2}\,\big]A_{N}
  \\[1.0ex]
  &=  -\eta^{MN}\widehat{\cal L}_{[\xi_1,\xi_2]_{C}}A_{N}  \ = \
  -\widehat{\cal L}_{[\xi_1,\xi_2]_{C}}A^{M}\;.
\end{split}
 \ee
We have verified explicitly that the
commutator~(\ref{liegencomm}) holds  acting on ${\cal H}^{MN}$.
With the identification of $\delta_\xi$ with $\widehat{\cal L}_\xi$
acting on ${\cal H}$ we have that up to terms that vanish because
of the strong form of the constraint,
 \be
 \label{closure}
  \big[\delta_{\xi_1},\delta_{\xi_2}\big]{\cal H}^{MN} =  \delta_{\xi_{12}}{\cal H}^{MN}\;,
 \ee
where $\xi_{12}=-[\xi_1,\xi_2]_{C}$.
The gauge transformations close according to the
C~bracket.
This is in agreement with \cite{Hull:2009zb} where it was shown that the algebra of the gauge transformations on $\cal E$ and $d$ is given by the C~bracket.

\subsection{Generalized Lie brackets and Dorfman brackets}

The usual Lie derivative of a vector field  defines the Lie bracket through $[X,Y]={\cal L}_XY$.
This suggests defining a generalized Lie bracket
through the generalized Lie derivative.  
This generalized Lie bracket,  which we will refer to as a D-bracket, is thus
defined by
\be
 \big[\,A,B\,\big]_{\rm D} \equiv \widehat{\cal L}_AB  \,.  
\ee
The D bracket is not skew-symmetric, as can be seen
using~(\ref{genLievecup}).
A short calculation shows that the D bracket differs from the  C bracket~(\ref{cbracketdef}) by a term which
 has the structure of a trivial gauge parameter:
 \bea
 \big[\,A,B\,\big]^{M}_{\rm D}
  \ = \ \big[\,A,B\,\big]^{M}_{\rm C}+\frac{1}{2}
  \partial^{M}\big(B^{N}A_{N}\big)\;.
 \eea
Generalized vectors that depend just on $x$ and not on $\tilde x$ decompose into a vector and a  1-form on
the usual $D$-dimensional space with coordinates $x^i$, and in that case it
was shown in~\cite{Hull:2009zb} that the C~bracket becomes precisely the Courant bracket.
In that same situation, the D-bracket becomes precisely the Dorfman bracket~(see, for example~\cite{Gualtieri}, section 3.2)
and our generalized Lie derivative becomes precisely the generalized Lie derivative introduced in~\cite{Grana:2008yw} leading to the standard transformations of the metric and $B$-field.
Our C~bracket, D~bracket and  generalized Lie derivative, however,  
 have the advantage of being $O(D,D)$ covariant. For any totally null $D$-dimensional subspace $N$ (i.e. any maximally isotropic subspace),
we showed in \cite{Hull:2009zb} that the C~bracket becomes the Courant bracket on $N$.
Similarly, the
D-bracket becomes the  Dorfman bracket on $N$ and the generalized Lie derivative becomes that of~\cite{Grana:2008yw} on $N$.

The D-brackets inherit the properties of the familiar
Dorfman bracket~\cite{Gualtieri}.\footnote{Since the Dorfman bracket
is not skew, it is usually not written as a bracket, but rather
as a product: $A \circ B$ denotes what we call $[A,B]_{\rm D}$.}
They are not skew:
 \bea
 \big[\,A,B\,\big]^{M}_{\rm D}+ \big[\,B,A\,\big]^{M}_{\rm D}
  \ =\
  \partial^{M}\big(B^{N}A_{N}\big)\;,
 \eea
 but their antisymmetrization gives the C bracket 
$ \big[\,A,B\,\big]_{\rm D}- \big[\,B,A\,\big]_{\rm D}   =
2 \big[\,A,B\,\big]_{\rm C}\;.$
It satisfies the Jacobi like identity 
\be
\big[A,\big[B,C\big]_{\rm D}\big]_{\rm D} = \big[\big[A,B\big]_{\rm D}\big],C\big]_{\rm D}
+ \big[B,\big[A,C\big]_{\rm D}\big]_{\rm D}\,.
\ee
Thus while the C bracket is anti-symmetric but does not satisfy the Jacobi identity, the D-bracket
is not anti-symmetric but does
satisfy a Jacobi identity.

\section{The gauge invariant action}\setcounter{equation}{0}
In this section we determine the gauge invariant action in terms of ${\cal H}^{MN}$, which is equivalent to
the original form (\ref{THEActionINTRO}). This action will be manifestly $O(D,D)$ invariant and is further
constrained by a discrete $\mathbb{Z}_2$ symmetry. Moreover, we construct a function ${\cal R}({\cal H},d)$ that transforms as
a gauge and $O(D,D)$ scalar and
show that the action, up to boundary terms, can be written in an Einstein-Hilbert-like form.

\subsection{The O(D,D) and gauge invariant action}\label{odinvact}

Given the $O(D,D)$ transformation properties of
${\cal H}^{MN}$,  the partial derivatives $\partial_N$, and  the metric
$\eta^{MN} = \eta_{MN}$, we can build $O(D,D)$ scalars
by simply contracting all indices consistently.  The $O(D,D)$ transformations are global and there is no complication whatsoever
with the derivatives.   Examples of  $O(D,D)$ scalars are
\be
\label{examples}
{\cal H}^{MN}\,\partial_{M}d\,
  \partial_{N}d \,, ~~     \partial^K {\cal H}^{MN} \partial_M{\cal H}_{KN}\,.
\ee
There are a number of such $O(D,D)$ scalars and what we are looking
for is a linear combination of them, each with two derivatives, that
is gauge invariant.  To simplify the problem we consider the one
additional discrete $\mathbb{Z}_2$ symmetry the action is supposed to
have.  This is the symmetry $b_{ij} \to - b_{ij}$ of the antisymmetric
tensor field.  This transformation must be accompanied by letting
$\tilde x \to - \tilde x$ as well as $\tilde \partial \to - \tilde \partial$.
In the original action~(\ref{THEActionINTRO}) this is the symmetry
under ${\cal E}_{ij} \to {\cal E}_{ji}$ as well as ${\cal D} \leftrightarrow
\bar {\cal D}$.  In our present notation, where
\be
\partial_M = \begin{pmatrix} \tilde\partial^i \\[1.0ex] \partial_i\end{pmatrix}\,,
\ee
we will write
\be
\partial_\bullet \to  Z\, \partial_\bullet \,, ~~~\hbox{with} ~~
Z= \begin{pmatrix} -1 & 0 \\ \phantom{-}0& 1 \end{pmatrix}\,,
\ee
where we used $\partial_\bullet$ to denote the column vector
associated with $\partial_M$.  The matrix $Z$ satisfies the simple
properties
\be
Z = Z^t = Z^{-1} \,,  ~~Z^2 = 1\,.
\ee
When $b_{ij} \to - b_{ij}$ the off-diagonal matrices in ${\cal H}^{MN}$
change sign. So do the off-diagonal matrices in ${\cal H}_{MN}$. This is accomplished by
\be
{\cal H}^{\bullet \bullet}  \to  Z {\cal H}^{\bullet \bullet} Z\,, ~~~~
{\cal H}_{\bullet \bullet}  \to  Z {\cal H}_{\bullet \bullet} Z \,.
\ee
The matrix $Z$ does not correspond to an $O(D,D)$ transformation.
Thus we find that
\be
\eta^{\bullet \bullet}  \not=  Z \,\eta ^{\bullet \bullet} Z\,, ~~~~
\eta_{\bullet \bullet}  \not=  Z \,\eta_{\bullet \bullet} Z \,.
\ee
We now see that terms built with $\partial_\bullet,  {\cal H}^{\bullet \bullet},$
and $ {\cal H}_{\bullet \bullet}$, with all indices contracted, will be $\mathbb{Z}_2$ invariant.  Indeed, each
index appears twice and, under the transformation, generate two
$Z$ matrices in a product  $ZZ =1$.  The $\mathbb{Z}_2$ invariance is violated if $\eta^{\bullet\bullet}$ or $\eta_{\bullet\bullet}$
are {\em needed} to write the term (${\cal H}_{\bullet\bullet}$ can be written with two $\eta$'s and ${\cal H}^{\bullet\bullet}$). Alternatively, the
$\mathbb{Z}_2$ invariance is violated if we {\em need} to use the derivatives
$\partial^M$ with an upper index.

The above $\mathbb{Z}_2$ constraint is quite strong.  It eliminates, for
example, the second term in~(\ref{examples}). In fact, one can convince
oneself that there is no $\mathbb{Z}_2$-invariant term with two derivatives
and two appearances of the generalized metric.  For terms that mix
the generalized metric and the dilaton there are four options:
\be
  ~\partial_{M}d\,\partial_{N}{\cal H}^{MN}\,, ~~{\cal H}^{MN}\,\partial_{M}d\,
  \partial_{N}d\,,
  ~~{\cal H}^{MN}\,\partial_{M}
  \partial_{N}d\,,~~  \partial_{M}\partial_{N}{\cal H}^{MN}\,.
\ee
The last one qualifies as an interaction because it is to be multiplied
by $e^{-2d}$, just as every other term.  By integration by parts we can
show that in an action the last two terms are simply linear combinations
of the first two. Thus our choices are
\be
 ~\partial_{M}d\,\partial_{N}{\cal H}^{MN}\,, ~~{\cal H}^{MN}\,\partial_{M}d\,
  \partial_{N}d\;.
\ee

Since we cannot have terms with just two generalized
metrics, we look for terms with three of them
(built without $\eta$).
There are just two options
\be
{\cal H}^{MN}\partial_{M}{\cal H}^{KL}
  \,\partial_{N}{\cal H}_{KL}\,, ~~~{\cal H}^{MN}\partial_{N}{\cal H}^{KL}\,\partial_{L}
  {\cal H}_{MK} \,.
  \ee
The action must be build by an appropriate linear combination of the
four terms listed above, and multiplied by $e^{-2d}$.
We claim that the gauge-invariant combination is
\be
\label{actionfirst}
S = \int dx \, d\tilde x  \, {\cal L}\,,
\ee
with
  \be
  \label{Haction}
  \begin{split}
 {\cal L} \ &= e^{-2d}\Big(\,\frac{1}{8}\,{\cal H}^{MN}\partial_{M}{\cal H}^{KL}
  \,\partial_{N}{\cal H}_{KL}-\frac{1}{2}
  \,{\cal H}^{MN}\partial_{N}{\cal H}^{KL}\,\partial_{L}
  {\cal H}_{MK} \\[1.0ex]
  &\hskip40pt
  -2\partial_{M}d\,\partial_{N}{\cal H}^{MN}+4{\cal H}^{MN}\,\partial_{M}d\,
  \partial_{N}d \Big)\,.
  \end{split}
 \ee
Rather than prove now the gauge invariance we first verify that the
above action  is equivalent  to the double field theory action in \cite{Hohm:2010jy}. Even more, the two corresponding Lagrangian densities are just identical.  As a check we perform a
 derivative expansion  ${\cal L}={\cal L}^{(0)}+{\cal L}^{(1)}+{\cal L}^{(2)}$ in 
$\tilde{\partial}$ 
as in \cite{Hohm:2010jy}. For
${\cal L}^{(0)}$ one finds
 \bea
  \begin{split}
   {\cal L}^{(0)} \ = \ e^{-2d}\Big(&\frac{1}{4}{\cal H}^{ij}\partial_{i}{\cal H}_{kl}
   \,\partial_{j}{\cal H}^{kl}+\frac{1}{4}{\cal H}^{ij}\partial_{i}{\cal H}_{k}{}^{l}\,
   \partial_{j}{\cal H}_{l}{}^{k}
   -\frac{1}{2}{\cal H}_{i}{}^{j}\partial_{j}{\cal H}_{k}{}^{l}\,\partial_{l}{\cal H}^{ik}
   -\frac{1}{2}{\cal H}_{i}{}^{j}\partial_{j}{\cal H}^{kl}\,\partial_{l}{\cal H}^{i}{}_{k}\\
   &-\frac{1}{2}{\cal H}^{ij}\partial_{j}{\cal H}_{k}{}^{l}\,\partial_{l}{\cal H}_{i}{}^{k}
   -\frac{1}{2}{\cal H}^{ij}\partial_{j}{\cal H}^{kl}\,\partial_{l}{\cal H}_{ik}
   -2\partial_{i}d\,\partial_{j}{\cal H}^{ij}+4{\cal H}^{ij}\,\partial_{i}d\,\partial_{j}d
   \Big)\;.
  \end{split}
 \eea
It is a straightforward though somewhat lengthy calculation to check that
 \be
   {\cal L}^{(0)} \ = \ e^{-2d}\Big(\frac{1}{4}g^{ij}\partial_{i}g_{kl}\,\partial_{j}g^{kl}
   -\frac{1}{2}g^{ij}\partial_{j}g^{kl}\,\partial_{l}g_{ik}-2\partial_{i}d\,\partial_{j}g^{ij}
   +4g^{ij}\partial_{i}d\,\partial_{j}d  -{1\over 12} H^2 \Big)\,,
 \ee  
where $H_{ijk} = \partial_i b_{jk} +\partial_j b_{ki}+\partial_k b_{ij}$.  This coincides with the expression found in eq.~(3.18) of~\cite{Hohm:2010jy}. Moreover, ${\cal L}^{(2)}$
turns out to be the `T-dual' expression, where we note that under inversion duality
 \be
  \partial_{i}\;\rightarrow\; \tilde{\partial}^{i}\;, \qquad
  {\cal H}^{ij}\; \rightarrow\; {\cal H}_{ij}\;, \qquad {\rm etc.}\;,
 \ee
i.e., ${\cal L}^{(2)}$ must also coincide with the corresponding expression in \cite{Hohm:2010jy}.
The lemma that two $O(D,D)$ scalars that
agree in one $O(D,D)$ frame are identical~\cite{Hohm:2010jy} shows that the two Lagrangians
are identical.
Thus (\ref{Haction}) is the correct rewriting of the Lagrangian in terms
of the generalized metric.

\bigskip
It is possible to understand the equality of the Lagrangians~(\ref{Haction})  and~(\ref{THEActionINTRO})
more directly. 
For this purpose it is useful
to define
 \be\label{gaugefixE}  
  {e}^{M}{}_{i} \ \equiv \  \begin{pmatrix} {\cal E}_{ji} \\[1.0ex] \delta^{j}{}_{i} \end{pmatrix}\;.
 \ee
This definition allows us to write
 \be\label{HMNa}
  {\cal H}^{MN} \ = \ {e}^{M}{}_{i}\,{e}^{N}{}_{j}\,g^{ij}-\eta^{MN}\;,
 \ee
as can be verified by a simple direct calculation of the components.
We also have
 \bea\label{calDbar}
  \ {e}^{M}{}_{i}\,\partial_{M} \ \equiv \ \bar{\cal D}_{i}\;,
 \eea
with the calligraphic derivative defined in (\ref{groihffkdf}). 
Next we will take terms in the new action
and write them in terms
of those in the old action.
For the last term in~(\ref{Haction}) the computation is rather simple:
 \be
 \begin{split}
  4{\cal H}^{MN}\partial_{M}d\,\partial_{N}d \ &=
  \ 4{e}^{M}{}_{i}{e}^{N}{}_{j}
  g^{ij}\partial_{M}d\,\partial_{N}d - 4\eta^{MN} \partial_{M}d\,\partial_{N}d \\[0.3ex]
  &=  \ 4
  g^{ij} {e}^{M}{}_{i}\partial_{M}d\, \, {e}^{N}{}_{j} \partial_{N}d \\[0.3ex]
  &=\ 4g^{ij}\bar{\cal D}_{i}d\,\bar{\cal D}_{j}d\;,
\end{split}
\ee
where we used the constraint, (\ref{HMNa}), and~(\ref{calDbar}).
The right-hand side is the last term in the
Lagrangian~(\ref{THEActionINTRO}).
Other terms require more work because they contain derivatives of ${\cal H}^{MN}$.  A short computation with the next to last term in~(\ref{Haction})
gives
  \be
  \begin{split}
  -2\,\partial_{M}d\,\partial_{N}{\cal H}^{MN} \ &= \
  -2\,\partial_{M}d~ \partial_N \bigl( {e}^M{}_{i} {e}^N{}_{j} g^{ij} \bigr) \\[0.5ex]
  &= \ -2 \,\tilde\partial^k d \, \bar {\cal D}_j {\cal E}_{ki} \, g^{ij}
  -2  \, \bar {\cal D}_i d \, \tilde\partial^k {\cal E}_{kj} \, g^{ij}  - 2\,
  \bar{\cal D}_i d \, \bar{\cal D}_j g^{ij} \,.
   \end{split}
 \ee
At this point we can replace the $\tilde\partial$-derivatives using the
identity $\tilde\partial^k = (\bar{\cal D}^k - {\cal D}^k)/2$, and one quickly
finds that
 \be
  -2\,\partial_{M}d\,\partial_{N}{\cal H}^{MN} \ = \
  g^{ij}g^{kl}\left({\cal D}_{l}d\,\bar{\cal D}_{j}{\cal E}_{ki}+\bar{\cal D}_{i}d\,
  {\cal D}_{l}{\cal E}_{kj}\right)\;.
 \ee
The right-hand side describes the next to last terms in the
Lagrangian~(\ref{THEActionINTRO})!  One must work harder to write
the pure generalized-metric terms in terms of ${\cal E}$ and calligraphic
derivatives.  But the results are still simple, with a rather direct
correspondence between the terms in the two actions.  We have
verified that
\be
\frac{1}{8}{\cal H}^{MN}\partial_{M}{\cal H}^{KL}
  \,\partial_{N}{\cal H}_{KL} =
  -\frac{1}{4} \,g^{ik}g^{jl}   \,   {\cal D}^{p}{\cal E}_{kl}\,
  {\cal D}_{p}{\cal E}_{ij}\,,
\ee
showing that the first terms in the Lagrangians are equal.  In doing this
computation the strategy is that terms with ${\cal E}$ fields and no derivatives have to combine and disappear, leaving at most metric
components $g_{ij}$. 
Finally, 
given the equality of the Lagrangians, the last two structures have to coincide,   
\be
-\frac{1}{2}{\cal H}^{MN}\partial_{N}{\cal H}^{KL}\,\partial_{L}
  {\cal H}_{MK} = \frac{1}{4}g^{kl} \bigl( {\cal D}^{j}{\cal E}_{ik}
  {\cal D}^{i}{\cal E}_{jl}  + \bar{\cal D}^{j}{\cal E}_{ki}\,
  \bar{\cal D}^{i}{\cal E}_{lj} \bigr)\,.
\ee
In total we conclude that all terms in the action can be identified naturally.

\bigskip
We note, in passing, that integrating by parts in the last term of the action defined by
(\ref{Haction}) we
can get the simpler, three term action:
\be\label{Haction99}
  S \ = \ \int dx d\tilde{x}\,e^{-2d}\Big(\frac{1}{8}{\cal H}^{MN}\partial_{M}{\cal H}^{KL}
  \,\partial_{N}{\cal H}_{KL}-\frac{1}{2}{\cal H}^{MN}\partial_{N}{\cal H}^{KL}\,\partial_{L}
  {\cal H}_{MK}
  + 2{\cal H}^{MN}\,\partial_{M}\partial_N d\,\Big)\,.
 \ee
Another set of integration by parts leads to an action where the Lagrangian
takes the form of $e^{-2d} {\cal R}$ where ${\cal R}$ is a gauge scalar.
This is what we discuss next.

\subsection{Generalized scalar curvature}

A reasonable assumption is that the analogue ${\cal R}$
of the scalar curvature is just
the dilaton equation of motion,
a feature that it shares with the scalar curvature
 constructed in~\cite{Hohm:2010jy},
  and that found in~\cite{Siegel:1993th}.
Using the Lagrangian~(\ref{Haction})
a simple computation gives the equation of motion of the dilaton, and
we thus define:
\be
 \label{simplerformR}
 \begin{split}
  {\cal R} \ \equiv  &~~~4\,{\cal H}^{MN}\partial_{M}\partial_{N}d
  -\partial_{M}\partial_{N}{\cal H}^{MN} \\[1.2ex]
   & -4\,{\cal H}^{MN}\partial_{M}d\,\partial_{N}d
   + 4 \partial_M {\cal H}^{MN}  \,\partial_Nd\;,\\[1.0ex]
    ~&+\frac{1}{8}\,{\cal H}^{MN}\partial_{M}{\cal H}^{KL}\,
  \partial_{N}{\cal H}_{KL}-\frac{1}{2}{\cal H}^{MN}\partial_{M}{\cal H}^{KL}\,
  \partial_{K}{\cal H}_{NL}\;.
 \end{split}
 \ee
The claim, to be proven in the following subsection, is that ${\cal R}$
so defined is a gauge scalar.
We can confirm that this is in fact the same as the
scalar in~\cite{Hohm:2010jy}.
The verification uses the equation of motion of the dilaton from
(\ref{Haction}).  The simplicity of this is that the Lagrangian here
reduces to ${\cal L}^{(0)}$.  The variation of the dilaton in the
Lagrangian (but not the exponential) then gives additional terms
that are total derivatives and a short computation shows that
they coincide with the total derivatives in equation~(C.27) of~\cite{Hohm:2010jy}.   This confirms
that the dilaton equation of motion does equal the curvature invariant.

We now confirm that the action (\ref{Haction}), up to total derivatives,
takes the form
\be
\label{actR}
S=  \int dx \, d\tilde x  \, e^{-2d} \, {\cal R}\;.
\ee
To see this we consider the last two terms in the Lagrangian of
(\ref{Haction}).  Simple manipulations show that
\be\label{Hactionv2}
\begin{split}
 & ~~~e^{-2d}\Big(-2\partial_{M}d\,\partial_{N}{\cal H}^{MN}+4{\cal H}^{MN}\,\partial_{M}d\,
  \partial_{N}d\Big) \\
  &= \partial_{M}(e^{-2d})\,\partial_{N}{\cal H}^{MN}+e^{-2d}\Big(
 -4{\cal H}^{MN}\,\partial_{M}d\,
  \partial_{N}d +  8{\cal H}^{MN}\,\partial_{M}d\,
  \partial_{N}d\Big)\\
    &= \partial_{M}(e^{-2d}\,\partial_{N}{\cal H}^{MN})
   +  e^{-2d}\Big( -\partial_{M}\partial_{N}{\cal H}^{MN}
    -4{\cal H}^{MN}\,\partial_{M}d\,
  \partial_{N}d \Big)-
   4{\cal H}^{MN}\,\partial_{M}( e^{-2d})\,
  \partial_{N}d\\
    &= \partial_{M}(e^{-2d}\,[ \partial_{N}{\cal H}^{MN}
     -4  {\cal H}^{MN}  \partial_{N}d])
   +  e^{-2d}\Big( \hskip-4pt-\partial_{M}\partial_{N}{\cal H}^{MN}
    -4{\cal H}^{MN}\partial_{M}d\,
  \partial_{N}d   + 4 \partial_M ({\cal H}^{MN} \partial_N d) \Big)\,.
 \end{split}
 \ee
We recognize that the terms within the last parentheses are
the first four terms in ${\cal R}$, as given in (\ref{simplerformR}).
Looking back at (\ref{Haction}) and ${\cal R}$ we conclude that
\be
{\cal L} = e^{-2d} {\cal
R} + \partial_{M}\Big(e^{-2d}\,[ \partial_{N}{\cal H}^{MN}
     -4  {\cal H}^{MN}  \partial_{N}d\,]\,\Big)\,.
\ee

\subsection{Proof of gauge invariance}

We prove now the gauge invariance of the action.  For this purpose
we consider (\ref{actR}) and we will show that  ${\cal R}$ is a gauge
scalar, namely,
 \be
 \label{gscalar}
  \delta_{\xi}{\cal R} \ = \widehat {\cal L}_\xi {\cal R} = \ \xi^{M}\partial_{M}{\cal R}\;.
 \ee
Given that the dilaton exponential transforms like a density:
\be
\label{dilp}
\delta_\xi \, e^{-2d} =  \partial_M (\xi^M e^{-2d})\,,
\ee
the invariance of the action $S$ follows immediately.

\bigskip
We use the same strategy as in \cite{Hohm:2010jy} to prove (\ref{gscalar}).  Since
all indices are properly contracted, we only need to focus on the non-covariant terms in the variation of partial derivatives. Thus, for example,
a short calculation shows that
 \be
 \label{parh}
  \delta_{\xi}\left(\partial_{M}{\cal H}^{KL}\right) \ = \ \widehat{\cal L}_{\xi}
  \left(\partial_{M}{\cal H}^{KL}\right)+\partial^{P}\xi_{M}\,\partial_{P}{\cal H}^{KL}
  -2\partial_{M}\partial_{P}\xi^{(K}\,{\cal H}^{L)P}+2\partial_{M}\partial^{(K}\xi_{P}\,
  {\cal H}^{L)P}\;.
 \ee
The first term is the covariant one. The second term vanishes due to the constraint, and consequently this term and analogous ones in the formulas below will be ignored in the following.  We will write, for any object $W$,
\be
\label{Delvar}
\delta_\xi  W =  \widehat{\cal L}_{\xi} W  + \Delta_\xi W\,,
\ee
 so that $\Delta_\xi W$ denotes the violation of $W$
to transform as the tensor associated with its index structure.
Since $\delta_\xi$
is a linear operation
\be
\begin{split}
\delta_\xi (W V) &=  (\delta_\xi W)  V  + W (\delta_\xi V)\\[0.4ex]
& =( \widehat{\cal L}_{\xi} W  + \Delta_\xi W)  V  + W ( \widehat{\cal L}_{\xi} V  + \Delta_\xi V)\\[0.5ex]
& = \widehat{\cal L}_{\xi} (W V)  + (\Delta_\xi W)  V  + W (  \Delta_\xi V)\,,
\end{split}
\ee
showing  that the violation $\Delta_\xi$ is also a derivation:
\be
\Delta_\xi (W V) =  (\Delta_\xi W)  V  + W (  \Delta_\xi V)\,.
\ee

\noindent
Using the notation in (\ref{Delvar}), the variation in (\ref{parh}) is
 \be
 \label{indexesup}
  \Delta_{\xi}\left(\partial_{M}{\cal H}^{KL}\right) \ = \
  -2\partial_{M}\partial_{P}\xi^{(K}\,{\cal H}^{L)P}+2\partial_{M}\partial^{(K}\xi_{P}\,
  {\cal H}^{L)P}\;.
 \ee
The contraction of the above is useful,
\bea
  \Delta_{\xi}\left(\partial_{M}{\cal H}^{MN}\right) \ = \
  -\partial_{P} (\partial \cdot \xi) \,{\cal H}^{PN}
   -\partial_{M}\partial_{P}\xi^{N}\,{\cal H}^{MP}
   +\partial_{M}\partial^{N}\xi_{P}\, {\cal H}^{MP}\;.
 \eea
Since $\delta_\xi \eta = \widehat {\cal L} \eta = 0$ we have $\Delta_\xi \eta =0$
and we can directly raise and lower indices  in formulae for $\Delta_\xi W$.  Thus,
(\ref{indexesup}) gives
  \bea
  \Delta_{\xi}\left(\partial_{K}{\cal H}_{NL}\right) \ = \ -2\partial_{K}\partial^{P}\xi_{(N}\,{\cal H}_{L)P}
  +2\partial_{K}\partial_{(N}\xi^{P}\,{\cal H}_{L)P}\;.
 \eea
We also need
 \be
 \begin{split}
  \Delta_{\xi}\left(\partial_{M}\partial_{N}{\cal H}^{MN}\right) &=~
  -2\,\partial_{M} (\partial\cdot \xi)\,\partial_{N}{\cal H}^{MN}
  -2\partial_{M}\partial_{N} (\partial \cdot \xi) \,{\cal H}^{MN}
  -\partial_{M}\partial_{N}\xi^{P}\,\partial_{P}{\cal H}^{MN}\;, \\[0.8ex]
   \Delta_{\xi}\left(\partial_{M}d\right)~ &=~
   -\frac{1}{2}\,\partial_{M}\, \partial \cdot \xi\;, \\[0.5ex]
  \Delta_{\xi}\left(\partial_{M}\partial_{N}d\right) &=~~
\partial_{M}\partial_{N}\xi^{P}\,\partial_{P}d-\frac{1}{2}\,\partial_{M}\partial_{N}
  (\partial \cdot \xi)\;.
  \end{split}
  \ee
In light of the above discussion, we need to show that
\be
\Delta_\xi {\cal R} = 0 \,.
\ee
We begin with the first two terms in ${\cal R}$, those that contain second
derivatives of fields.  A short calculation
shows that
\be
\begin{split}
\Delta_\xi \Bigl( 4\,{\cal H}^{MN}\partial_{M}\partial_{N}d
  -\partial_{M}\partial_{N}{\cal H}^{MN}\Bigr)
  =&~  4 {\cal H}^{MN} \,\partial_M \partial_N \xi^P \partial_P d
  + 2 \partial_M \partial\cdot \xi  \, \partial_N {\cal H}^{MN}\\
&   + \partial_M \partial_N \xi^P  \partial_P {\cal H}^{MN}\,.
  \end{split}
\ee
The virtue of the above combination of terms is that variations with
three derivatives on $\xi$ cancelled out.   Next we aim to cancel
the term above of the form ${\cal H}\,\partial^2 \xi\, \partial d$.
For this we use the next two terms in ${\cal R}$.  A short computation
gives
\be
\Delta_\xi \Bigl( 4 \partial_M {\cal H}^{MN} \partial_N d - 4{\cal H}^{MN} \,\partial_M d \partial_N d \Bigr)
= - 4 {\cal H}^{MN} \,\partial_M \partial_N \xi^P \partial_P d
\,- \,2 \partial_M \partial\cdot \xi  \, \partial_N {\cal H}^{MN}\,.
\ee
Comparing with the previous equation we see that two terms are
in fact cancelled and we get
\be
\label{fldkf}
\Delta_\xi \Bigl( 4\,{\cal H}^{MN}\partial_{M}\partial_{N}d
  -\partial_{M}\partial_{N}{\cal H}^{MN}
  +4 \partial_M {\cal H}^{MN} \partial_N d - 4{\cal H}^{MN} \,\partial_M d \partial_N d\Bigr) = \partial_M \partial_N \xi^P  \partial_P {\cal H}^{MN}\,.
\ee
The violation in the right-hand side can be cancelled by one of the
remaining terms in ${\cal R}$:
\be
\label{fifthv}
\begin{split}
\Delta_\xi \Bigl( -\frac{1}{2}{\cal H}^{MN}\partial_{M}{\cal H}^{KL}\,
  \partial_{K}{\cal H}_{NL}\Bigr) &=~- \partial_M \partial_N \xi^P  \partial_P {\cal H}^{MN}\\[0.5ex]
  &~~~~~+ \partial_K {\cal H}^{MN} \,\partial_M \bigl(\underline{ \partial^L \xi_P} - \partial_P \xi^L\bigr) {\cal H}^{KP} {\cal H}_{NL}\;.
\end{split}
\ee
We can show that the underlined term is in fact zero because
it is equal to minus itself:
 \be\label{strange}
 \begin{split}
 \partial_K {\cal H}^{MN}  {\cal H}^{KP} {\cal H}_{NL}   \,\partial_M  \partial^L \xi_P =&  -\partial_{K}{\cal H}_{NL}{\cal H}^{KP}{\cal H}^{MN}\,
  \partial_{M}\partial^{L}\xi_{P} \\
  =& -\partial_{K}{\cal H}^{NL} {\cal H}^{KP}{\cal H}_{MN}\,
  \partial^{M}\partial_{L}\xi_{P}
  \\   =& -\partial_{K}{\cal H}^{NM}{\cal H}^{KP}{\cal H}_{LN}\,
  \partial^{L}\partial_{M}\xi_{P}
  \\
  =& -\partial_{K}{\cal H}^{MN}{\cal H}^{KP}{\cal H}_{NL}\,
 \partial_{M} \partial^{L}\xi_{P} \;.
\end{split}
\ee
In the first step we used  $ \partial_K {\cal H}^{MN}   {\cal H}_{NL}
= -  {\cal H}^{MN}   \partial_K {\cal H}_{NL}$, which follows from
${\cal H}^{MN}   {\cal H}_{NL}= \delta^M_L$.   As a result (\ref{fifthv}) becomes
\be
\label{fifthv99}
\begin{split}
\Delta_\xi \Bigl( -\frac{1}{2}{\cal H}^{MN}\partial_{M}{\cal H}^{KL}\,
  \partial_{K}{\cal H}_{NL}\Bigr) &=~- \partial_M \partial_N \xi^P  \partial_P {\cal H}^{MN}- \partial_K {\cal H}^{MN} \, {\cal H}^{KP} {\cal H}_{NL} \,
  \partial_M \partial_P \xi^L  \;.
\end{split}
\ee
The first term on the right-hand side is suitable to cancel the violation
in (\ref{fldkf}) but we got an additional term.  This term requires
the consideration of the one remaining term in ${\cal R}$.  A short
calculation gives
\be
\label{lvgvm}
\Delta_\xi \Bigl( {1\over 8} {\cal H}^{MN} \partial_M {\cal H}^{KL}
\partial_N {\cal H}_{KL}\Bigr)
= {1\over 2} \partial_M {\cal H}^{KL} \,\partial_N \bigl( \partial_K \xi^P
- \partial^P \xi_K\bigr)  {\cal H}^{MN} {\cal H}_{LP}\;.
\ee
Using manipulations similar to those in (\ref{strange})  show that the
two terms in the above right-hand side are actually equal so that
\be
\label{lvgsg}
\begin{split}
\Delta_\xi \Bigl( {1\over 8} {\cal H}^{MN} \partial_M {\cal H}^{KL}
\partial_N {\cal H}_{KL}\Bigr)
=& ~ \partial_M {\cal H}^{KL} \,
 {\cal H}^{MN} {\cal H}_{LP} \partial_N  \partial_K \xi^P\,,\\
 =& ~\partial_K {\cal H}^{MN} \, {\cal H}^{KP} {\cal H}_{NL} \,
  \partial_M \partial_P \xi^L\,,
 \end{split}
\ee
using additional manipulations for the last step.
This result, together with (\ref{fifthv99}), gives
\be
\label{lvgjn}
\Delta_\xi \Bigl({1\over 8} {\cal H}^{MN} \partial_M {\cal H}^{KL}
\partial_N {\cal H}_{KL} -\frac{1}{2}{\cal H}^{MN}\partial_{M}{\cal H}^{KL}\,
  \partial_{K}{\cal H}_{NL}\Bigr)=
  - \partial_M \partial_N \xi^P  \partial_P {\cal H}^{MN}\,.
\ee
At this point it is clear that the $\Delta_\xi$ violation of these
two terms cancels precisely with the violation of the other four
terms in ${\cal R}$, as shown in (\ref{fldkf}).  We thus find that
$\Delta_\xi {\cal R}=0$, which is what we wanted to show.
This completes the proof of gauge invariance of the action.

\subsection{Generalized Ricci curvature}\label{genricci}

We have seen that the dilaton field equation gives a generalized scalar 
${\cal R}$ that can be viewed as a generalisation of the 
scalar curvature.
In this subsection we consider the field equation for ${\cal H}$ which provides a natural generalization of the Ricci curvature.
The change in the action (\ref{Hactionx}) under a general variation $\delta {\cal H}^{MN}$ of $  {\cal H}^{MN}$ is
\be
\label{varact}
\delta S\ = \ \int dx d\tilde{x}\,e^{-2d}~ \delta {\cal H}^{MN} 
{\cal K}_{MN}   \;, 
\ee
where
\be
\label{kis}
\begin{split}  
{\cal K} _{MN}\equiv 
~&~
\frac{1}{8}\, \partial_{M}{\cal H}^{KL}
  \,\partial_{N}{\cal H}_{KL}
  -{1\over 4} (\partial_L - 2 (\partial_L d) ) 
  ({\cal H}^{LK} \partial_K {\cal H}_{MN})
  +2 \,\partial_{M}\partial_N d\,  
    \\[1.0ex]
  & \hskip-10pt -\frac{1}{2} \partial_{(M}{\cal H}^{KL}\,\partial_{L}
  {\cal H}_{N)K}
  + {1\over 2} (\partial_L - 2 (\partial_L d) )  \bigl({\cal H}^{KL} \partial_{(M}
   {\cal H}_{N)K}
  + {\cal H}^K{}_{(M}  \partial_K {\cal H}^L{}_{N)}  \bigr) \,.
   \end{split}
\ee
As ${\cal H}$ is constrained to satisfy 
${\cal H}\eta{\cal H} =  \eta^{-1}$,  
 the equations of motion are found by
 considering variations that preserve this constraint.
 The varied field ${\cal H}'= {\cal H} + \delta {\cal H} $ will satisfy $  {\cal H}'\eta{\cal H}' \ = \ \eta^{-1}$  provided
 \be
\delta {\cal H}\,\eta{\cal H}  + {\cal H}\eta \, \delta{\cal H} \ = \ 0\,.
 \ee
Using~(\ref{sasmatrix}) we 
rewrite the above as 
  \be
\delta {\cal H}\,S^t  + S \, \delta{\cal H} \ = \ 0\,,
 \ee
 and recalling that $S^2 =1$ we have the constraint
 \be
 \label{eomcons}
 \delta {\cal H}\  = \ -  S \, \delta{\cal H} \, S^t \,.
 \ee
Since $\half (1\pm S)$,
 acting on vectors $V=V^M$ with upper indices, 
 can be viewed as projectors into subspaces with 
$S$  eigenvalues $\pm 1$, any matrix 
 $M=M^{MN}$ can be
 viewed as a bivector and so 
  written as the
sum of four projections into independent subspaces: 
\be
\begin{split}
M \ = \ & ~~ \frac 1 4 (1+S) {M} (1+S^t) + \frac 1 4 (1+S) {  M} (1-S^t)\\[1.4ex]
&\hskip-7pt +  \frac 1 4 (1-S) {M} (1+S^t) + \frac 1 4 (1-S) {  M} (1-S^t)\,.
\end{split}
\ee
 It then follows that the general solution of~(\ref{eomcons}) is 
 given by  
 \be
\delta {\cal H}= \frac 1 4 (1+S)\,  {\cal M} (1-S^t) + \frac 1 4 (1-S)  {\cal M} (1+S^t)\,,
 \ee
where $ {\cal M}$ is an arbitrary matrix that must be symmetric to 
guarantee that $\delta {\cal H}$ is symmetric.
Inserting this in~(\ref{varact})  
and letting ${\cal K}$ denote the matrix
with components ${\cal K}_{MN}$ gives
\be
\begin{split}
\delta S \ =\ & \int dx d\tilde x\, e^{-2d} \,\hbox{Tr} \bigl( \,\delta {\cal H}\, {\cal K}\bigr) \\[1.0ex]
=\ & \int dx d\tilde x\, e^{-2d} \,\hbox{Tr} \Bigl( \,{\cal M}\, \Bigl[
{1\over 4} (1-S^t)\, {\cal K}\, (1+ S) 
+ {1\over 4} (1+S^t)\, {\cal K}\, (1-S)\Bigr] \Bigr)\,.
\end{split}
\ee
The field equation is then 
\be
{\cal R}_{MN}\ =\ 0\,,
\ee
where the matrix ${\cal R}$, whose components are
${\cal R}_{MN}$, is given by 
\be
\label{riss}
{\cal R} \equiv  \frac 1 4 (1-S^t) {\cal K} (1+S) 
+ \frac 1 4 (1+S^t) {\cal K} (1-S)
\,.
\ee
Restoring the indices we have
\be
\label{riss_index}
{\cal R}_{MN} \equiv  \frac 1 4 (\delta_M{}^P -S^P{}_M) \,{\cal K}_{PQ}\, (\delta^Q{}_N +S^Q{}_N) + \frac 1 4 (\delta_M{}^P +S^P{}_M) 
\,{\cal K}_{PQ} \,(\delta^Q{}_N -S^Q{}_N)
\,.
\ee
The field equation ${\cal R}_{MN}=0$ combines the field equations of the metric $g$ and the $b$-field in an $O(D,D)$ covariant form and ${\cal R}_{MN}$ provides a generalized Ricci curvature.
We will discus it further in section 5.4.

\section{Vielbein formulations}\setcounter{equation}{0}

In this section we discuss reformulations of the double field theory written in terms of vielbeins or coset variables instead of the metric $\cal H$, introducing variables similar to those used in dimensional reduction of supergravity theories.  
When dimensionally reduced on a $D$-torus, the familiar field theory of gravity plus b-field arising in supergravity theories gives 
a theory with $O(D,D)$ duality symmetry. In particular, the
scalar fields originating from
the metric and b-field on the torus take values in the coset space
$O(D,D)/O(D)\times O(D)$ \cite{Maharana:1992my,Meissner:1991zj,Kleinschmidt:2004dy}.
Scalar fields 
such as these  that parameterize  
a coset space $G/H$ 
can be represented by a 
group-valued field ${\cal V}(x)\in G$, which
depends only on the non-compact coordinates 
$x^\alpha$ and transforms as 
\be\label{cosetsym}
  {\cal V}^{\prime}(x) \ = \ g\,{\cal V}(x)\,h(x)\;, \qquad
 g\in G \;, \quad h(x)\in H\;
 \ee
 under local $H$ and rigid $G$ transformations.
 The theory can also be written in terms of
 ${\cal H}= {\cal V}^{\dagger}  {\cal V}$, which is $H$-invariant and reduces to ${\cal H}= {\cal V}^{t}  {\cal V}$
if $\cal V$ is  a real matrix, as in our case.
 
    For $G/H=O(D,D)/O(D)\times O(D)$,
 the matrix
${\cal V}^M{}_{A}$ has an $O(D,D)$ index $M$ and
a $O(D)\times O(D)$ index $A$. The vielbein ${\cal V}^M{}_{A}$ corresponds to the matrix
$h_{\cal E}$ in (\ref{representative}), which indeed transforms from
the left by the $G= O(D,D)$ action and is well-defined only up to local
$H= O(D)\times O(D)$ transformations from the right.
We define the $H$-invariant ${\cal H}^{MN}$~by
 \be\label{cosetH}
  {\cal H}^{MN} \ \equiv \ {\cal V}^{M}{}_{A}\,{\cal V}^{N}{}_{B}\,\delta^{AB}\;.
 \ee
This definition coincides with (\ref{smnvm}) and thus
${\cal H}^{MN}$ 
is the inverse  generalized metric.
Any action for which ${\cal V}$ enters only through ${\cal H}$ is
automatically $H$-invariant.
The standard sigma model Lagrangian
reads ${\cal L}={\rm Tr}\big[ {\cal H}^{-1}\partial^{\alpha}{\cal H}\,{\cal H}^{-1}\partial_{\alpha}{\cal H}\big]$.
An alternative formulation of this sigma model was found by
Maharana and Schwarz~\cite{Maharana:1992my} involving a vielbein $e^M{}_{a}$, where $a=1,\ldots,D$, and in this formulation the local symmetry is $GL(D,\R)$ instead of
$O(D)\times O(D)$.

Siegel generalised this  by using similar variables for the whole space-time, not just an internal torus. In \cite{Siegel:1993xq} he rewrote the metric and b-fields in $D$-dimensional flat space
in terms of a Maharana-Schwarz-like vielbein
${e }^M{}_{a}$ 
depending on {\it all}
 the space-time coordinates,
 not just the non-compact ones.
Then in \cite{Siegel:1993th} he extended this further  to
a doubled space-time with $2D$ coordinates $X$ transforming as a vector under $O(D,D)$ and a vielbein $e^M{}_{A}(X)$ with a local $GL(D,\R)\times GL(D,\R)$ symmetry. This formulation
reduces to the 
coset space formulation with one gauge choice and
to the Maharana-Schwarz-like formulation with another.
Neither $e^M{}_{A}(X)$ or $e^M{}_{a}(X)$ are coset representatives, but
fixing the  $GL(D,\R)\times GL(D,\R)$ symmetry to $O(D)\times O(D)$ does give a representative  of the coset $O(D,D)/O(D)\times O(D)$.
Models  with fields taking values in a coset $G/H$ and  depending on coordinates $X$ that transform under $G$
were discussed in \cite{Hillmann:2009ci}, motivated by earlier use of such variables in e.g.~\cite{Borisov:1974bn} and~\cite{West:2000ga}. A key feature is that such models allow for more general $G$-invariant actions, as
the derivatives $\partial _M$ now carry the same kind of index as ${\cal H}^{MN}$ or ${\cal V}^M{}_{A}$. Consequently, contractions between indices on derivatives and indices on matrices are now possible, as arising in our action (\ref{Haction}).

Here we will discuss reformulations of our theory in terms of vielbeins
$e^M{}_A(X)$ or $e^M{}_a(X)$, and use these to explore the geometry further.
In this way we show how our formalism is related to 
that of Siegel,  giving a 
different approach to his formalism here.

\subsection{General frames}

We start by choosing a basis of vector fields $e^{M}{}_{A}$ for the doubled space,
where $M$ is the usual vector index and $A=1,\ldots,2D$ labels the basis. Then
$e^{M}{}_{A}$  is a $2D\times 2D$ {\em invertible} matrix field, and
its inverse  $e^A{}_M$
 can be regarded as a vielbein for the doubled space.
 The metrics $\cal H$ and $\eta $ then have frame components
 \bea\label{doubleH}
  {\cal H}_{AB} \ \equiv \ e^{M}{}_{A}\,e^{N}{}_{B}\,{\cal H}_{MN}\;
 \eea
and
 \bea\label{flatmetric}
 \hat{\eta}_{AB} \ \equiv \ e^{M}{}_{A}\,e^{N}{}_{B}\,\eta_{MN}\;,
  \eea
where the $\,\hat{}\,$ indicates that for general frames this will be a function of $X$.
There is a local $GL(2D,\R)$ action on frames
\be
e^{M}{}_{A}\;\to\; e^{M}{}_B\, \Lambda ^B{}_A\;,~~~
\Lambda(X) \in GL(2D,\R)\,.
\ee

The inverse metrics  have frame components defined by
 \bea\label{doubleup}
  {\cal H}^{AB} \ \equiv \ e^{A}{}_{M}\,e^{B}{}_{N}\,{\cal H}^{MN}\;, \qquad
  \hat{\eta}^{AB} \ \equiv \ e^{A}{}_{M}\,e^{B}{}_{N}\,\eta^{MN}\;,
  \eea
  and it follows from these definitions and (\ref{inverseH}) that, as expected, these are the inverses of $ {\cal H}_{AB}$ and $\hat{\eta}_{AB}$:
\be
 \label{inverseHup}
 {\cal H}^{AC}{\cal H}_{CB}\ = \ \delta ^A{}_B\;,
  \qquad
 \hat{\eta}^{AC}\hat{\eta}_{CB}\ = \ \delta ^A{}_B
 \;.
 \ee
Using the above formulae, it follows that  
 \be
  e^{A}{}_{M} ( \eta^{MN} \,e^{B}{}_{N}\, \hat{\eta}_{BC}) \ = \ \delta ^A{}_C\;,
  \ee
  so that the inverse vielbein is given by
   \be
e^M{}_C \ = \  \eta^{MN} \,e^{B}{}_{N}\, \hat{\eta}_{BC}\;.
  \ee
  Thus it is fully consistent to raise and lower  $M,N, \ldots$ indices
  with $\eta _{MN}$ and the tangent space indices
  $A,B,\ldots$ with $\hat{\eta}_{AB}$, and we will do so throughout this section.

The tangent space group
$GL(2D,\R)$ can be reduced by restricting to a basis with special properties.
For example, a basis that is orthonormal with respect to
$\eta_{MN}$ will have  
 \be
e^M{}_A e^N{}_B \, \eta_{MN} 
= \ \begin{pmatrix}
  0 & 1 \\ 1 & 0 \end{pmatrix}  \quad \to \quad
    \hat{\eta}_{AB} \ = \ \begin{pmatrix}
  0 & 1 \\ 1 & 0 \end{pmatrix}\;.
 \ee
Restricting to such bases will restrict the tangent space group to $O(D,D)$.
Similarly, a basis that is orthonormal with respect to
${\cal H}_{MN}$ will have  
 \be
e^M{}_A e^N{}_B \, {\cal H}_{MN} 
= \delta_{AB} \quad \to \quad
   {\cal H}_{AB} \ = \ \delta_{AB}\;.
 \ee
Restricting to such bases will restrict the tangent space group to $O(2D)$.\footnote{This is for the case in which $g_{ij}$ and ${\cal H}_{MN}$ are positive definite. For $g_{ij}$ of
signature $(p,q)$, ${\cal H}_{MN}$ has signature $(2p,2q)$ and orthonormal frames will have
 ${\cal H}_{AB}$ the constant Minkowski-type metric of signature $(2p,2q)$, and the tangent space group would be $O(2p,2q)$. Throughout this section, we will present results for the case in which
 $g_{ij}$ is positive definite, but our formulae all have natural generalisations to the case of general signature.}
Restricting to frames which are orthonormal for both  metrics $\cal H$ and $\eta $
reduces the  tangent space group to $O(D)\times O(D)$.

\subsection{Frames with $GL(D,\R)\times GL(D,\R)$ symmetry}

Here we will be interested in a different reduction of the frame bundle in which the structure group is reduced to
$GL(D,\mathbb{R})\times GL(D,\mathbb{R})$.
The doubled space is equipped with the two metrics
$\cal H$ and $\eta $ so that $S=\eta^{-1} {\cal H}$ satisfies $S^2=1$ and is an almost local product structure or almost real structure (the analogue of an almost complex structure satisfying $J^2=-1$).
This allows the splitting $T=T_+\oplus T_-$ of the tangent bundle $T$ of the doubled space
into the subbundle $T_+$ of vectors with $S$ eigenvalue $+1$ and the subbundle $T_-$  with $S$ eigenvalue $-1$.
We will choose a basis of $D$ vectors
$e^M{}_a$ for $T_-$ ($a = 1, \ldots, D$) and a basis $e^M{}_{\bar a}$  ($\bar a = 1, \ldots, D$) for $T_+$.
Then $A=(\bar{a},a)$ is an composite index  and we have
 \bea\label{ecomp}
  e^{M}{}_{A} \ = \ \begin{pmatrix}
  e^{M}{}_{\bar{a}} & e^{M}{}_{a} \end{pmatrix}
  \ = \ \begin{pmatrix} e_{i\bar{a}} & e_{ia} \\[0.5ex]
  e^{i}{}_{\bar{a}} & e^{i}{}_{a} \end{pmatrix}
  \;,
 \eea
 where  
 \be
 \label{sactione}
 \begin{split}
  S e_a \ &= \ - e_a, \\
   Se_{\bar a} \ &= \ \phantom{-} e_{\bar a}\;.
   \end{split}
 \ee
As ${\cal H}_{MN} = \eta_{MP} S^P_{~~N}$  
these imply
\be
\label{frameev}
\begin{split}
{\cal H}_{MN}\, e^N{}_a \ &= \ -\eta_{MN}\, e^N{}_a \, , \\[0.5ex]
 {\cal H}_{MN}\, e^N{}_{\bar a} \ &= \ ~\eta_{MN}\, e^N{}_{\bar a}\;.
\end{split}
\ee
Contracting each of these with 
a vielbein we obtain
\be
\begin{split}
{\cal H}_{MN}\, e^{M}{}_{a}\, e^N{}_b 
\ &= \ -\eta_{MN}\, e^{M}{}_{a} \,e^N{}_b \, \quad \to \quad
{\cal H}_{ab} =  - \hat\eta_{ab} \,, \\[1.0ex]
{\cal H}_{MN}\, e^{M}{}_{\bar{a}}\,e^N{}_{\bar b} \ &= \ 
~\,\,\eta_{MN}\,e^{M}{}_{\bar{a}}\, e^N{}_{\bar b}\;\quad \to \quad
{\cal H}_{\bar a\bar b} = ~\,\,  \hat\eta_{\bar a\bar b}  \,.
\end{split}
\ee
Additional information comes by considering ${\cal H}_{MN}\, e^{M}{}_{{a}}\,e^N{}_{\bar b}$.  Evaluating this term using the first and second equations
in~(\ref{frameev}) gives
\be\label{offdiagonal}
\begin{split}
{\cal H}_{MN}\, e^{M}{}_{{a}}\,e^N{}_{\bar b} \ = \
({\cal H}_{NM}\, e^{M}{}_{{a}})\,e^N{}_{\bar b}
\ &= \ -\eta_{MN}\,e^{M}{}_{{a}}\, e^N{}_{\bar b} \,, \\[0.5ex]
{\cal H}_{MN}\, e^{M}{}_{{a}}\,e^N{}_{\bar b}  \ = \
({\cal H}_{MN}\, \,e^N{}_{\bar b}) e^{M}{}_{{a}}
\ &= \ ~ ~\eta_{MN}\,e^{M}{}_{{a}}\, e^N{}_{\bar b} \,.\\
\end{split}
\ee
Since the two evaluations differ by a sign, the term in question
vanishes. This means that  
\be
\begin{split}
{\cal H}_{a\bar b } =  0\,, & ~~~ \hat\eta_{a\bar b} = 0\,, \\[0.5ex]
 {\cal H}_{\bar a b } =  0 \,, 
&~~~\hat\eta_{\bar a b} = 0\,.
\end{split}
\ee
We define $g_{ab}$ and $g_{\bar a \bar b}$ as the nonvanishing
components of the flattened metric $\hat \eta_{AB}$, 
with factors of two introduced for later convenience,  
\be
\label{giss}
\begin{split}
g_{ab}\ \equiv& \ 
- \frac 12\, {\eta}_{MN}\, e^{M}{}_{a}\, e^N{}_b 
= -{1\over 2} \,\hat\eta_{ab}\;, \\[1.5ex]
g_{\bar a \bar b}\ \equiv& \ ~~~\frac 12  \,{\eta}_{MN}\, e^{M}{}_{\bar{a}}\,e^N{}_{\bar b}\;  = ~~{1\over 2} \,\hat\eta_{\bar a\bar b}\,.
\end{split}
\ee
The above results are summarized
by giving the  flat components of ${\cal H}$ and $\eta$:
\be
\label{metaftcons}
  {\cal H}_{AB} = 2\begin{pmatrix}
    {g}_{\bar a \bar b}&  0\\[0.5ex]
    0  &
      {g}_{ab}  \end{pmatrix}\;, \qquad
  \hat{\eta}_{AB} =2 \begin{pmatrix}
    {  g}_{\bar a \bar b}&  0\\[0.5ex]
    0  &
     - {  g}_{ab}  \end{pmatrix}\,.
\ee
If the original metric $g_{ij}$ is positive definite, then $  {\cal H}_{MN} $ and
$  {\cal H}_{AB} $ are positive definite. Thus, ${g}_{ab} $ and $ {g}_{\bar a \bar b}$ are positive definite as well, while $\hat{\eta}_{AB}$ has signature $(D,D)$. (For other signatures, if  $g_{ij}$  is invertible, then so are ${g}_{ab} $ and $ {g}_{\bar a \bar b}$.)

Choosing frames in this way reduces the tangent space group to
$GL(D,\mathbb{R})\times GL(D,\mathbb{R})$
with one $GL(D,\mathbb{R})$ factor acting on the indices $a,b$ as frame rotations of $T_-$ and
the other  $GL(D,\mathbb{R})$ factor acting on the indices $\bar a, \bar b$ as frame rotations of $T_+$.
 The $O(D,D)$ and the local tangent space symmetries then act on the frame field as follows
 \be\label{bige}
   {e'}^{M}{}_{A}{}
  (X^{\prime}) \ = \  h ^{M}{}_{N}\,
   \,e^{N}{}_{B}(X)  \,\Lambda^B{}_{A}(X) \,\;,
  \ee
with $h\in O(D,D)$ and $\Lambda(X)\in GL(D,\mathbb{R})\times GL(D,\mathbb{R})$,
so that
\be
\Lambda^A{}_{B} \ = \ \begin{pmatrix}
    {  \Lambda}^{\bar a}{}_{ \bar b}&  0\\[0.5ex]
    0  &
      {  \Lambda}^a{}_{b}  \end{pmatrix}\;,
\ee
with $ {  \Lambda}^{\bar a}{}_{ \bar b}$ in the first  $GL(D,\mathbb{R})$
and
$   {  \Lambda}^a{}_{b}$ in the second  $GL(D,\mathbb{R})$. Here $X'=hX$ as before, so that the coordinates transform under $O(D,D)$ but are inert under the tangent space group.
This is Siegel's vielbein formalism with $ GL(D,\mathbb{R})\times GL(D,\mathbb{R})$ symmetry  \cite{Siegel:1993th}.

Following~\cite{Siegel:1993th} we consider gauge fixing the tangent space symmetry.
We use one $GL(D,\R)$ symmetry to choose
$g_{ab}=\frac 12 \delta _{ab}$ and the other
$GL(D,\R)$ symmetry to choose
$g_{\bar a \bar b}=\frac 12
\delta _{\bar a \bar b}$. Then
 \bea\label{gaugefixingg}
   {\cal H}_{AB} \ = \  \begin{pmatrix} \delta_{\bar{a}\bar{b}} & 0 \\
   0 & \delta_{ab} \end{pmatrix} \ \equiv \ \delta_{AB} \;, \qquad
  \hat{\eta}_{AB} \ = \  \begin{pmatrix} \delta_{\bar{a}\bar{b}} & 0 \\
   0 &-  \delta_{ab} \end{pmatrix}\;.
 \eea
The basis is then orthonormal with respect to both $\cal H$ and $\eta$ and
the tangent space group is reduced to
$O(D)\times O(D)$.
Then (\ref{doubleH}) implies
\bea\label{doubleHH}
  {\cal H}_{MN} \ = \   {\delta}_{AB}\, e^A{}_M\,e^B{}_N  \;,
 \eea
so that
${\cal H}^{-1}= e^te$  
and $e^A{}_M $ is a vielbein
for the generalized metric.
The matrix $\hat{\eta}_{AB}$ appearing in (\ref{gaugefixingg}) differs by a similarity transformation from $\eta_{MN}$:
 \be
 \label{vmgg}
  \sigma \, \hat{\eta} \,\sigma^t \ =\eta\, \;,  \qquad
  \sigma \ = \ \frac{1}{\sqrt{2}} \begin{pmatrix} 1 & -1 \\ 1 & \phantom{-}1 \end{pmatrix}
  \ = \ (\sigma^{t})^{-1}\;.
 \ee
We then have  
 \bea
 \label{werow}
  \eta_{AB} \ = \ \hat{e}^{M}{}_{A}\,\hat{e}^{N}{}_{B}\,\eta_{MN}\;, \quad
  \hbox{where}\quad 
  \hat{e}^{M}{}_{A} \ \equiv \ e^{M}{}_{B}\,\sigma_{A}{}^{B}\;.
 \eea
This is verified by expanding the right-hand side, using~(\ref{flatmetric}),
and the relation~(\ref{vmgg}) that follows from the gauge fixing.
The
 result (\ref{werow}) means that $\hat{e}$ is an $O(D,D)$ group element.
Thus,
$\hat{e}^M{}_A$ is an $O(D,D)$ matrix transforming under  a rigid $O(D,D)$ transformation $h$ and
a local $O(D)\times O(D)$ transformation $\Lambda (X) $ as
\be
\hat{e}'(X') \ =  \ h\, \hat{e}(X)\, \Lambda (X)\;,
\ee
where $X'=hX$.  The gauge  equivalence classes of $\hat{e}$ under the local $O(D)\times O(D)$ symmetry can then be identified with fields taking values in  the coset space
$O(D,D)/O(D)\times O(D)$.
In this way
we recover the familiar  coset space variables.

\subsection{Frames with $GL(D,\R)$ symmetry}

We now return to the general situation with the full  $ GL(D,\mathbb{R})\times GL(D,\mathbb{R})$ symmetry and show that  the geometry can be formulated in terms of the frames
 for $T_+$:  
\be
  e^{M}{}_{\bar a} \ = \ \begin{pmatrix} e_{i\bar{a}}  \\[0.5ex]
  e^{i}{}_{\bar{a}}  \end{pmatrix}\;.
 \ee
There is a local $GL(D,\R)$ symmetry acting on the index $\bar a$.
In matrix notation,
\be\label{GLD}
e^M\,\rightarrow\, e^M \,\Lambda\,,  ~~~\Lambda\in GL(D,\mathbb{R})\;.
\ee 
We note that (\ref{metaftcons}) implies
\be
  {\cal H}^{AB} \ = \ \frac 1 2 \begin{pmatrix}
    {g}^{\bar a \bar b}&  0\\[0.5ex]
    0  &
      {g}^{ab}  \end{pmatrix}\;, \qquad
  \hat{\eta}^{AB} \ = \ \frac 1 2 \begin{pmatrix}
    {  g}^{\bar a \bar b}&  0\\[0.5ex]
    0  &
     - {  g}^{ab}  \end{pmatrix}\;,
\ee
where ${g}^{ab}$ is the inverse of ${g}_{ab}$ and ${g}^{\bar a \bar  b}$ is the inverse of ${g}_{\bar  a \bar  b}$.
Then
\be
\label{metaftconsb}
  {\cal H}^{AB}  \ = \ \begin{pmatrix}
    {g}^{\bar a \bar b}&  0\\[0.5ex]
    0  &
     0  \end{pmatrix}- \hat{\eta}^{AB}\,.
\ee
Acting on this with $e^M{}_A\,e^N{}_B$ gives
\be\label{HMN}
  {\cal H}^{MN} \ = \ {e}^{M}{}_{\bar a}\,{e}^{N}{}_{\bar b}\,g^{\bar a \bar b}-\eta^{MN}\;.
 \ee
As ${g}_{\bar a \bar b}$  is given in terms of the $ e^{M}{}_{\bar a}$ by (\ref{giss}), it follows that
(\ref{HMN}) gives an expression for the generalised metric in terms of the $T_+$ frames $ e^{M}{}_{\bar a}$ alone.

\medskip
The components of $ {\cal H}^{MN}$ are given in (\ref{H}) and this can be used to find expressions for $g_{ij}$ and $b_{ij}$ in terms of the frame fields.
First, the lower right block of (\ref{H}) is ${\cal H}^{ij} = g^{ij}$ and using this in (\ref{HMN}) gives
\be\label{rootg}
g^{ij} \ = \ e^i{}_{\bar a}\, e^j{}_{\bar b}\,{g}^{\bar a \bar b}\;.
\ee
Remarkably, this implies that $e^i{}_{\bar a}$ is non-degenerate and can be viewed as $D\times D$ frame fields to convert the indices $i,j,\ldots$ to flat indices $\bar a, \bar b ,\ldots$.  
Moreover,  $g_{\bar a \bar b}$ are precisely the frame components of $g_{ij}$.
The  inverse of $e^i{}_{\bar a}$
is then the vielbein
\be
e^{\bar a}{}_i \ = \ g_{ij}\,
e^j{}_{\bar b}\,{g}^{\bar a \bar b}\;.
\ee
Similarly, the upper right block of (\ref{H}) is ${\cal H}_i{}^{j} =b_{ik} g^{kj}$, and using this in (\ref{HMN}) gives
\be
b_{ik} g^{kj} \ = 
\ e_{i \bar a}\, e^j{}_{\bar b}\,{g}^{\bar a \bar b}-\delta_i{}^j\;.
\ee
Multiplication by $g_{jp}$ quickly leads to  
\be\label{Efac}
{\cal E}_{ij}\ = \ g_{ij}+b_{ij} \ = \ e_{i \bar a}\, e^{\bar a}{}_j\;. 
\ee
Thus, in addition to the vielbein
$e^{\bar a}{}_i $ that gives a \lq square root' of the metric $g_{ij}$ in (\ref{rootg}), there is a
field $e_{i \bar a}$ which can be viewed as a
second vielbein that  incorporates the $b$-field and which gives a factorisation of $\cal E$.
The contraction of upper and lower $\bar a$ indices in (\ref{Efac}) implies that
${\cal E}_{ij}$ is invariant under $GL(D,\R)$. 
Thus the generalised metric ${\cal H}$ and $\cal E$ are both given in terms of the frame fields 
$e^M{}_{\bar a}$  ($\bar a = 1, \ldots, D$) for $T_+$.
This is  essentially  the two-vielbein formalism 
given in~\cite{Siegel:1993xq} for fields depending on the spacetime coordinates and extended to doubled fields in~\cite{Siegel:1993th}.

The linear transformation of $e^M{}_{\bar a}$ under
$O(D,D)$ then implies that ${\cal E}_{ij}=e_{i \bar a}\, e^{\bar a}{}_j$ transforms by the fractional linear transformation (\ref{ODDtrans}).
To see this we use matrix notation and denote the $D\times D$ matrix components of $e^{M}{}_{\bar a}$  
 as $e$ and $\tilde{e}$, such that  
 ${\cal E}=e\,\tilde{e}^{-1}$. They transform under the $O(D,D)$ group element in (\ref{ODDtrans}) as follows  
 \bea
  \begin{pmatrix} e \\[0.5ex] \tilde{e}  \end{pmatrix}\;\rightarrow\;  \begin{pmatrix} a & b\\[0.5ex] c & d  \end{pmatrix}
  \begin{pmatrix} e \\[0.5ex] \tilde{e}  \end{pmatrix} \ = \ \begin{pmatrix} a e+b\tilde{e} \\[0.5ex] c e+d\tilde{e}  \end{pmatrix}\;.
 \eea
This implies for the transformation of ${\cal E}$
 \begin{eqnarray}
  {\cal E} &\rightarrow 
  & \left(ae+b\tilde{e}\right)\left(ce+d\tilde{e}\right)^{-1} 
  \ = \ \left(ae\tilde{e}^{-1}+b\right)
  \tilde{e}\tilde{e}^{-1}\left(ce\tilde{e}^{-1}+d\right)^{-1} \\ \nonumber
  &=& \left(a{\cal E}+b\right)\left(c{\cal E}+d\right)^{-1}\;,
 \end{eqnarray}
which is the 
 fractional linear transformation~(\ref{ODDtrans}), as we wanted to show.

We next fix the local $GL(D,\R)$ symmetry (\ref{GLD}).
One possibility  
is the gauge choice $g_{\bar a \bar b}=\delta _{\bar a \bar b}$. This makes the frame orthonormal so that one has the usual 
$g_{ij}= e^{\bar a } {}_i\, e^{\bar b} {}_j\, \delta _{\bar a \bar b}$
and  the tangent space group is reduced to $O(D)$.
Alternatively, the $GL(D,\R)$ symmetry can be completely fixed by choosing the gauge
\be\label{GLDfixing}
GL(D,\mathbb{R}) ~\hbox{gauge fixing}:~~ e^{i}{}_{\bar a} \ = \ \delta^{i}{}_{\bar a}\,.
\ee
In this gauge  we identify
flat indices ${\bar a,\bar b}$ and world indices $i,j$.  It follows that
$g_{ij}$ and $g_{\bar a\bar b}$ become identical matrices on  
account of~(\ref{rootg}).  Moreover, we have
 $e^{\bar a}_{~i}=\delta^{\bar a}{}_i$  and 
equation~(\ref{Efac}) gives 
 \be
\label{eise}
{\cal E}_{ij}\ = \ e_{i \bar a}\, e^{\bar a}{}_j \ = \ e_{i j}\,.
\ee
As a result, we have 
\be
\label{hsdgfkjrdhjg}
  e^{M}{}_{i}
  \ = \  \begin{pmatrix} {\cal E}_{ji} \\[1.0ex] \delta^{j}{}_{i} \end{pmatrix}\;. 
\ee
This is then precisely  the field defined in  (\ref{gaugefixE}).
The derivatives with frame indices  reduce in this gauge as follows 
 \bea\label{calDbar99}
D_{\bar a} \, \equiv\, e^M{}_{\bar a}\,\partial_M \quad \Rightarrow \quad 
D_{{i}} \ =  \ e^{M}{}_{{i}}\,\partial_{M} \ = \ \bar{\cal D}_{i}\;,  
 \eea
recovering the calligraphic derivative as in (\ref{calDbar}). In this way we 
provide  a geometric setting for the equations that were
used in \S\ref{odinvact} in order to discuss the equivalence of the actions in terms of ${\cal E}$ and ${\cal H}$.

\bigskip
Similar arguments lead to
completely analogous results for the frames of $T_-$.
The frames
of $T_-$ are
\bea\label{ecompv}
  e^{M}{}_{  a} \ = \ \begin{pmatrix} e_{i {a}}  \\[0.5ex]
  e^{i}{}_{ {a}}  \end{pmatrix}\;.
 \eea
The generalised metric is given by
\be\label{HMNb}
  {\cal H}^{MN} \ = \ {e}^{M}{}_{  a}\,{e}^{N}{}_{  b}\,g^{  a   b}+\eta^{MN}\;,
 \ee
the inverse metric is
\be
g^{ij} \ = \ e^i{}_{  a}\, e^j{}_{  b}\,{g}^{  a   b}\;,
\ee
and $e^i{}_{  a}$ is non-degenerate with    inverse
given by the vielbein
\be
e^{  a}{}_i  \ = \ g_{ij}\,
e^j{}_{  b}\,{g}^{  a   b}\;.
\ee
Furthermore, 
\be
{\cal E}_{ji} \ = \ g_{ij}-b_{ij} \ = \ -e_{i   a}\, e^{  a}{}_j\;,
\ee
so that
\be
{\cal E}_{ij} \ = \ -e_{j  a}\, e^{  a} {}_i\;. 
\ee

The $GL(D,\R)$ symmetry acting on the indices $a,b,\ldots$  can be completely fixed by choosing the gauge
$ e^{i}{}_{  a}=\delta^{i}{}_{  a}$.
In this gauge  we identify
flat indices ${  a,  b}$ and world indices $i,j$,  and the matrices $g_{ij}$
and $g_{ab}$. 
The frames $e^M{}_a$ become
 \be
  e^M{}_i   
   \ = \  \begin{pmatrix} 
   -{\cal E}_{ij} \\[1.0ex] \delta^{j}{}_{i} \end{pmatrix} \;.
 \ee
The flattened derivatives in this gauge 
are
 \bea\label{calDbar999}
 D _{ a}\ = \ e^M{}_{  a}\,\partial_M \quad \Rightarrow \quad 
 D_i \ = \   e^{M}{}_{i}\,\partial_{M} \ \equiv \  {\cal D}_{i}\;,  
 \eea
giving the unbarred calligraphic derivative defined in (\ref{groihffkdf}).

\subsection{Gauge choices and applications}

The doubled space has two metrics: the fixed metric $\eta$ 
that appears in the constraint
$\eta^{MN} \partial _M \partial_N =0$
and the dynamical metric $\cal H$, which encodes the space-time metric and $b$-field.
As we have seen,   
the geometry could instead be formulated in terms of $\eta$ and the frames $e^M{}_{\bar a}$,
 in terms of $\eta$  and the frames $e^M{}_{a}$,
or  in terms of $e^M{}_A$ and $\eta$.

  The theory can be rewritten in terms of the frame field $e^M{}_{\bar a}$, giving  a 
  theory with a
  local $GL(D,\R)$ symmetry.
  An advantage of this formalism is that  $e^M{}_{\bar a}$ is 
  unconstrained.\footnote{   
  If we  are to 
  restrict to non-degenerate
  metrics $g_{ij}$, then certain   
 invertibility requirements need to be imposed on $e^M{}_{\bar a}$.}
In this subsection we  use the  frame formulation 
to investigate the relation between the formulation in terms of 
${\cal E}_{ij}$ and in terms of ${\cal H}$.

As we have seen,  the $GL(D,\R)$ gauge symmetry can be completely fixed by
  the gauge choice $e^{i}{}_{\bar a} \ = \ \delta^{i}{}_{\bar a}$
so that the frame field is given in terms of $\cal E$ by (\ref{hsdgfkjrdhjg}).
 Together with~(\ref{HMN})     
 this gives a rewriting of the ${\cal H}$-based theory in terms of $\cal E$, and this was the strategy used in section 4 to show the equivalence of the action (\ref{Hactionx}) in terms of $\cal H$ and the action (\ref{THEActionINTRO}) in terms of $\cal E$.

Next  we  use the frame formalism  
to relate the gauge transformations of ${\cal E}$ to those of ${\cal H}$.
  The gauge and $GL(D,\mathbb{R})$
transformations of the frame field
${ e}^{M}{}_{\bar a}$ are
\be\label{gaugeE}
\begin{split}
  \delta{ e}^{M}{}_{\bar a} \ &= \ \widehat{\cal L}_{\xi}{ e}^{M}{}_{\bar a}+
  { e}^{M}{}_{\bar b}\, \Sigma^{\bar b}{}_{\bar a}
  \\[1.0ex]
   \ &= \
  \xi^{K}\partial_{K}{ e}^{M}{}_{\bar a}-\partial_{K}\xi^{M}\,{ e}^{K}{}_{\bar a}
  +\partial^{M}\xi_{K}\,{ e}^{K}{}_{\bar a}
  +{ e}^{M}{}_{\bar b}\, \Sigma^{\bar b}{}_{\bar a}\;,
 \end{split}
 \ee
where $\Sigma^{\bar b}{}_{\bar a}$ is the local $GL(D,\mathbb{R})$ parameter. Then
${\cal H}^{MN}$ is given by~(\ref{HMN}) and is a $GL(D,\mathbb{R})$ singlet
and transforms with the  generalized Lie derivative
under the $\xi^M$ gauge transformations.
The gauge condition  $e^{i}{}_{\bar a}=\delta^{i}{}_{\bar a}$ is not preserved by the
$\xi$ gauge transformations and so these must be accompanied by compensating $GL(D,\mathbb{R})$ transformations.
The transformation of the gauge-fixed component $ { e}^{i}{}_{\bar{a}}$ is
\begin{eqnarray}
  \delta { e}^{i}{}_{\bar{a}} &=& -\partial_{K}\xi^{i}\,{ e}^{K}{}_{\bar{a}}
  +\tilde{\partial}^{i}\xi_{K}\,{ e}^{K}{}_{\bar{a}}
  +{ e}^{i}{}_{\bar{b}}\,\Sigma^{\bar{b}}{}_{\bar{a}} 
  \\
  \nonumber
  &=& -\bar{\cal D}_{\bar{a}}\xi^{i}+\tilde{\partial}^{i}\tilde{\xi}_{\bar{a}}
  +\tilde{\partial}^{i}\xi^{k}\,{ e}_{k\bar{a}}+
  \delta^{i}{}_{\bar{b}}\,\Sigma^{\bar{b}}{}_{\bar{a}}
  \;,
 \end{eqnarray}
 where we used  (\ref{calDbar99}), which holds after gauge-fixing.
 In order to preserve  the gauge condition  
  we need  $\delta e^{i}{}_{\bar{a}}=0$ and therefore  
 a $\xi$-transformation must be accompanied by a compensating $GL(D,\R)$ transformation with parameter
 \bea\label{Sigmafix}
  \Sigma^{j}{}_{i} \ = \ \bar{\cal D}_{i}\xi^{j}-\tilde{\partial}^{j}\tilde{\xi}_{i}
  -\tilde{\partial}^{j}\xi^{k}\,{ \cal E}_{ki}\;,
 \eea
where we have used that after gauge-fixing ${\cal E}_{ij}=e_{i j}$, and `world indices' 
are identified with `flat indices'.
Now, from (\ref{hsdgfkjrdhjg})
we have in this gauge
\be
\label{hsdgfkjrdhjgsdf}
  \delta e^{M}{}_{i}
  \ = \  \begin{pmatrix} \delta {\cal E}_{ji} \\[1.0ex] 0 \end{pmatrix}\;.
\ee
The  $\xi $ variation of ${\cal E}_{ij}$ can thus be found by substituting
(\ref{hsdgfkjrdhjgsdf}) in
 (\ref{gaugeE}) and using~(\ref{Sigmafix}),  
 \begin{eqnarray}
  \delta{\cal E}_{ij} &=& \xi^{K}\partial_{K}{\cal E}_{ij}-\partial_{K}\tilde{\xi}_{i}\,
  {e}^{K}{}_{j}   
   +\partial_{i}\xi_{K}\,{e}^{K}{}_{j}+{\cal E}_{ik}\,\Sigma^{k}{}_{j}
  \\ \nonumber
  &=& \xi^{K}\partial_{K}{\cal E}_{ij}-\bar{\cal D}_{j}\tilde{\xi}_{i}
  +\partial_{i}\tilde{\xi}_{j}+\partial_{i}\xi^{k}\,{\cal E}_{kj}+
  \left(\bar{\cal D}_{j}\xi^{k}-\tilde{\partial}^{k}\tilde{\xi}_{j}
  -\tilde{\partial}^{k}\xi^{p}\,{\cal E}_{pj}\right){\cal E}_{ik}\\ \nonumber
  &=& {\cal D}_{i}\tilde{\xi}_{j}
  -\bar{\cal D}_{j}\tilde{\xi}_{i}+\xi^{K}\partial_{K}{\cal E}_{ij}+
  {\cal D}_{i}\xi^{k}\,{\cal E}_{kj}
  +\bar{\cal D}_{j}\xi^{k}\,{\cal E}_{ik}\;.
 \end{eqnarray}
 This is precisely the gauge
transformation~(\ref{finalgtINTRO}) of ${\cal E}_{ij}$, and thus we have shown that this can be understood as arising from a geometric transformation and a compensating tangent space rotation.

Alternatively, the theory can be rewritten in terms of the frame field $e^M{}_A$, giving a formulation with local $GL(D,\R)\times GL(D,\R)$ symmetry as in \cite{Siegel:1993th}.
This can be done by writing  ${\cal H} + \eta $ in terms of   $e^M{}_{\bar a}$ using (\ref{HMN}) and
writing  ${\cal H} - \eta $ in terms of   $e^M{}_{a}$ using (\ref{HMNb}).
Note that (\ref{offdiagonal}) implies
 \bea\label{constraint}
    e^{i}{}_{a}\,e_{i\bar{b}} + e_{ia}\,e^{i}{}_{\bar{b}} \ = \ 0\;. 
 \eea
The gauge transformations are 
 \bea
  \delta e^{M}{}_{A} \ = \ \widehat{\cal L}_{\xi}\,e^{M}{}_{A}
  +e^{M}{}_{B}\Sigma^{B}{}_{A}\, \;,  
 \eea
where the parameter $\Sigma$ takes values in the Lie algebra
$\frak{gl}(D,\mathbb{R})\oplus\frak{gl}(D,\mathbb{R})$.

We can gauge-fix the $GL(D,\mathbb{R})\times GL(D,\mathbb{R})$ completely
by setting $e^{i}{}_{\bar{a}}=\delta^{i}{}_{\bar{a}}$ and $e^{i}{}_{a}=\delta^{i}{}_{a}$.
Then the indices $a,b...$ and $\bar a, \bar b....$ are both identified with the world indices $i,j,...$ and
   we identify the component $e_{i\bar{a}}$ with ${\cal E}_{ij}$ as before. Then the constraint
(\ref{constraint}) determines $e_{ia}$ to be
$e_{ia}=-{\cal E}_{ai}$; thus we have
 \bea\label{squareE}
  e^{M}{}_{A} \ = \ \begin{pmatrix} {\cal E}_{i\bar{a}} & -{\cal E}_{ai} \\[0.5ex]
  \delta^{i}{}_{\bar{a}} & \,\delta^{i}{}_{a} \end{pmatrix}\;.
 \eea

 The components of the `flattened' derivative in this gauge become, on identifying the indices $i$ with both $a$ and $\bar a$,
 \bea\label{flatder}  
  D_{A} \ \equiv \ e^{M}{}_{A}\,\partial_{M} \qquad \Rightarrow\qquad
 D_{a} \ = \ {\cal D}_{a} \;, \quad D_{\bar{a}} \
  = \ \bar{\cal D}_{\bar{a}} \;.
 \eea
In this gauge the flattened derivatives are the calligraphic derivatives as in
(\ref{calDbar99}) and (\ref{calDbar999}).

As an illustration  of this formalism we translate the 
strong 
constraint
$\partial^{M}f\partial_{M}g=0$  
  into the language of calligraphic derivatives.
It follows from the second equation in~(\ref{doubleup}) that
\be
\eta^{MN} = \hat \eta^{AB}  \, e^M{}_{A} e^N{}_{B}.
\ee 
It thus follows that
 \bea
 \begin{split}
  0 \ &= ~\eta^{MN} \partial_{M}f\,\partial_{N}g\   
  = \ \partial^{M}f\,\partial_{M}g \ = \
  \hat{\eta}^{AB}\,e^{M}{}_{A}\,e^{N}{}_{B}\,\partial_{M}f\,\partial_{N}g \ = \
  \hat{\eta}^{AB}\,D_{A}f\,D_{B}g \\[0.3ex]
  \ &= \
  \frac{1}{2} g^{\bar a\bar b} \,D_{\bar a} f\,D_{\bar b}g
  -{1\over 2} g^{ab} D_af\,D_{b}g \ = \ \frac{1}{2} g^{ij} \,\bar{\cal D}_{i} f\,
  \bar{\cal D}_{j}g
  -{1\over 2} g^{ij} {\cal D}_if\,{\cal D}_{j}g   \\[0.3ex]
  \ &=\ -\frac{1}{2}\left({\cal D}^{i}f\,{\cal D}_{i}g-\bar{\cal D}^{i}f\,\bar{\cal D}_{i}g\right)\;,
 \end{split}
 \eea
where we used the expression for $\hat \eta^{AB}$ in (\ref{metaftconsb}), 
 the identification of $g_{ab}$ and $g_{\bar a\bar b}$ with $g_{ij}$,
and (\ref{flatder}). 
This is the constraint in calligraphic
derivatives \cite{Hohm:2010jy}.

\bigskip

The frame fields are useful in the discussion of the generalized Ricci curvature introduced in~\S\ref{genricci}.  The tensor ${\cal K}_{MN} $ given in (\ref{kis}) has frame components
\be
{\cal  K}_{AB}\ = \ {\cal K}_{MN} e^M{}_Ae^N{}_B \ = \   
\begin{pmatrix}
  {\cal K}_{\bar a \bar b}&   {\cal K}_{\bar a b}\\[0.5ex]
     {\cal K}_{a \bar b}  &
      {\cal K}_{ab}  \end{pmatrix}\;.
\ee
As $\frac 1 2(1+S) $ projects onto barred indices and 
$\frac 1 2(1-S) $ projects onto unbarred
 indices (see~(\ref{sactione})),  the frame components of (\ref{riss}) are
\be
{\cal R}_{AB}={\cal R}_{MN} e^M{}_Ae^N{}_B=  
\begin{pmatrix}
 0&   {\cal K}_{\bar a b}\\[0.5ex]
     {\cal K}_{a \bar b}  &
  0 \end{pmatrix} \,,
\ee
so that the unmixed components vanish, $ {\cal R}_{a b} = {\cal R}_{\bar a \bar b} =0$, and the mixed ones are determined by the mixed components of ${\cal K}$, so that
$ {\cal R}_{a \bar b} = {\cal K}_{a \bar b} $.

\section{Conclusions and Outlook}

In this paper we have
reformulated  the background independent double field theory of~\cite{Hohm:2010jy}
in terms of the generalized metric ${\cal H}_{MN}$.
The action and gauge transformations
 simplify significantly
  when written in terms of ${\cal H}$, and the proof of gauge invariance is considerably easier than the one given in~\cite{Hohm:2010jy}.
  The generalized metric  transforms covariantly
under $O(D,D)$ and as a result the action and gauge transformations are manifestly $O(D,D)$ covariant.
 The gauge symmetry acts nonlinearly on the fields $g_{ij}$ and $b_{ij}$
(or ${\cal E}_{ij}$) used for the formulation in~\cite{Hohm:2010jy} but
becomes linear when written in terms of
${\cal H}_{MN}$.  
The gauge algebra of
double field theory is characterized by a C bracket that is the natural 
$O(D,D)$ covariant
extension of the Courant bracket to doubled fields. 
The C bracket reduces
to the Courant bracket when the fields are
restricted to a null subspace. 
The action in terms of ${\cal H}$ can be seen as a
 generalization of a non-linear sigma model based on the coset space
$O(D,D)/(O(D)\times O(D))$
 in which the coordinates transform under $O(D,D)$.
The generalized metric 
${\cal H}_{MN}$ can be viewed as a composite field defined in terms of a metric $g$
and an antisymmetric tensor $b$.  
Alternatively,   ${\cal H}_{MN}$ can be viewed
as an elementary field that is constrained to be a symmetric
$O(D,D)$ matrix. 
 The constraint  can be solved by writing ${\cal H}$ 
in terms of frame fields, so
that these frame fields could   
be viewed as the basic fields of the theory.

We defined a  generalized Lie derivative that was suggested
by the linear form of the gauge transformations of ${\cal H}^{MN}$
and introduced  generalized tensors that transform with this derivative. 
We explored the properties of these derivatives in some detail.
It is crucial that  
 the Lie derivative of the $O(D,D)$ metric
$\eta^{MN}$ vanishes, so that the $O(D,D)$ structure is preserved by the gauge transformations.   The commutator of two generalized
Lie derivatives is again a generalized Lie derivative with parameter obtained through the
C~bracket.  The generalized scalar curvature ${\cal R}$, 
built with two derivatives acting on the generalized metric and the dilaton,  
 indeed transforms as a generalized scalar.

We have discussed the relation of our work to that of Siegel~\cite{Siegel:1993th,Siegel:1993xq}.
The frame fields with  
tangent space symmetry $GL(D,\mathbb{R})\times GL(D,\mathbb{R})$
have simple transformation properties and the fields ${\cal H}$  and ${\cal E}$ were constructed in terms of these.   
The frame variables were useful in showing the relation between the theory written
in terms of ${\cal H}$ and that written  in terms of ${\cal E}$.
 In~\cite{Siegel:1993th}, Siegel introduced  covariant derivatives and  curvatures constructed from the vielbein fields and used these to write an action.  Its relation to our actions should be investigated.    
 
  The \lq generalized metric'  ${\cal H}_{MN}$ should properly be regarded as a {\em conventional} metric on the doubled space.  It is only unusual in  that this metric  is constrained to be an $O(D,D)$ matrix.  
The metric ${\cal H}_{MN}$
is the natural extension of the  generalized metric of generalized geometry to doubled fields.
It is intriguing that two key ingredients of generalized geometry play central roles in the double field theory: Courant brackets and the
generalized metric.  
There is much that remains to be understood of the geometry underlying the double field theory.
We have found a natural field strength transforming as a scalar under the gauge transformations, and
the field equation for ${\cal H}$ gives a generalisation of the Ricci tensor, 
but we do not have an understanding of these as  curvatures.
It would be of considerable interest to develop a geometric understanding of our results, perhaps combining ideas from generalized geometry with the  constructions of Siegel.
 This would help in constructing  
  gauge-invariant higher derivative actions.

Our results in this paper use the strong version of the $\partial_M\partial^M =0$ constraint, which requires that  
all fields and products of fields are in the kernel of $\partial_M\partial^M$.
This strong form of the constraint implies that all fields and parameters depend on just $D$ of the $2D$ coordinates, so that the theory can be viewed as a conventional theory living on a $D$ dimensional subspace of the doubled spacetime.
The most important outstanding question is whether there is a gauge invariant theory in which only the weak form of the constraint is imposed,  so that each field satisfies the constraint, but products of fields need not do so. Such a theory   
would depend non-trivially on all the coordinates of the doubled spacetime and so would be a true double field theory. 
This theory   was constructed to cubic order in~\cite{Hull:2009mi} and shown to be gauge invariant using only the weak form of the constraint.
Its extension to higher orders, however, necessarily involves new structures 
and the  explicit appearance of a  
projector onto the kernel of 
$\partial_M\partial^M$~\cite{Hull:2009mi}.     
We hope that the geometric structures discussed in this paper will be useful in the quest for such a theory.

\subsection*{Acknowledgments}
We acknowledge helpful discussions
with Ashoke Sen, Warren Siegel,
and Dan Waldram.
This work is supported by the U.S. Department of Energy under 
research agreement DE-FG02-05ER41360. The work of OH is supported by the DFG -- The German Science Foundation.

\end{document}